\documentclass[preprint,superscriptaddress,nofootinbib]{revtex4-1}
\usepackage{graphicx}
\usepackage{amsmath,amssymb,color,colortbl}
\usepackage{cancel}
\usepackage{comment}
\def\beq#1\eeq{\begin{equation}#1\end{equation}}
\def\bal#1\eal{\begin{align}#1\end{align}}

\newcommand{\llww}{$\ell^\pm \ell^{\prime\pm} W^\mp W^\mp$}

\begin{document}

\preprint{OU-HET-1042}
\preprint{KANAZAWA-20-02}

\title{
Probing charged lepton number violation \\ via $\ell^\pm \ell^\prime{}^\pm W^\mp W^\mp$
}

\author{Mayumi Aoki}
\email{mayumi@hep.s.kanazawa-u.ac.jp}
\affiliation{
Institute for Theoretical Physics,
Kanazawa University,
Kanazawa 920-1192, Japan
}
\author{Kazuki Enomoto}
\email{kenomoto@het.phys.sci.osaka-u.ac.jp}
\affiliation{
Department of Physics,
Osaka University,
Toyonaka,
Osaka 560-0043, Japan
}
\author{Shinya Kanemura}
\email{kanemu@het.phys.sci.osaka-u.ac.jp}
\affiliation{
Department of Physics,
Osaka University,
Toyonaka,
Osaka 560-0043, Japan
}

\begin{abstract}
 We study impacts of dimension-five lepton-number violating operators associated with two same-sign weak bosons, $\ell^\pm \ell^{\prime \pm} W^\mp W^\mp$, on current and future experiments for neutrino oscillation, lepton-number violating rare processes and  high-energy collider experiments.
These operators can contain important information on the origin of tiny neutrino masses, which is independent of that from the so-called Weinberg operator. 
We examine constraints on the coefficients of the operators by the neutrino oscillation data.
Upper bounds on the coefficients are also investigated by using the data for processes of lepton number violation such as neutrinoless double beta decays and $\mu^-$-$e^+$ conversion. 
These operators can also be directly tested by searching for lepton-number violating dilepton production via the same-sign W boson fusion process at high-energy hadron colliders like the Large Hadron Collider.
We find that these operators can be considerably probed by these current and future experiments.
\end{abstract}

\maketitle

\section{Introduction}

In 2012, the Higgs boson was discovered at the LHC~\cite{cite:Higgs_Discovery}, and the existence of all particles predicted in the Standard Model (SM) was confirmed empirically.
On the other hand, the SM cannot explain some observed phenomena, such as baryon asymmetry of the Universe~\cite{Sakharov:1967dj}, the existence of dark matter~\cite{Aghanim:2018eyx} and neutrino oscillation~\cite{cite:Neutrino_oscillation}.
It is one of the important goals of current particle physics to establish the theory beyond the SM which can explain the origin of these mysterious phenomena.

The observed neutrino oscillation indicates that neutrinos have small but non-zero masses. 
This smallness would suggest that the origin of small neutrino masses is different from the electroweak symmetry breaking. 
It would be natural to consider that neutrinos have Majorana-type masses, instead of Dirac-type masses.
In this case, the theory beyond the SM is expected to have a source of Lepton Number Violation (LNV) at high energies, which provides the origin of tiny Majorana-type masses of neutrinos at low energies.

Such a high-scale physics may be well described by Effective Field Theories (EFTs) with the electroweak gauge symmetry.
One of the most important operators of LNV is so-called the Weinberg operator~\cite{Weinberg:1979sa}, which is a dimension-five operator.
There are many models where small Majorana masses of neutrinos are generated via the Weinberg operator, like the type-I~\cite{ref:seesaw,Schechter:1980gr}, the type-II~\cite{Schechter:1980gr,ref:HTM}, the type-III seesaw mechanisms~\cite{Foot:1988aq}, and models where neutrino masses are radiatively generated at one-loop~\cite{Zee:1980ai,ref:Ma}, two-loop~\cite{Zee:1985id,Babu:1988ki} and three-loop level~\cite{ref:KNT,ref:AKS}.
  
If the lepton number is not conserved at high energies, we generally have various higher-dimensional operators of LNV~\cite{cite:Wudka,Gustafsson:2014vpa,Lehman:2014jma,Babu:2001ex,deGouvea:2007qla,Angel:2012ug,Herrero-Garcia:2019czj}, in addition to the Weinberg operator.
After the electroweak symmetry breaking, some of them yield the dimension-five charged LNV operators $\ell^{\pm}\ell^{\prime \pm} W^\mp W^\mp$ where $\ell(\ell^\prime)$ represents a charged lepton $e$, $\mu$ or $\tau$, and $W^\pm$ are the weak bosons. 
Electroweak gauge invariant origins of these dimension-five operators are dimension-seven (dimension-nine) operators in the case that leptons in the operators are left-handed (right-handed)~\cite{cite:Wudka,Gustafsson:2014vpa}.
In general, their coefficients are independent of that of the Weinberg operator, and can be related to neutrino masses~\cite{delAguila:2011gr,Gustafsson:2012vj,Gustafsson:2014vpa}.

There are many low-energy experiments searching for LNV phenomena,
such as neutrinoless double beta decays ($0\nu\beta\beta$)~\cite{Umehara:2008ru,KamLAND-Zen:2016pfg,Arnold:2016qyg,Aalseth:2017btx,Alduino:2017ehq,Albert:2017owj,Agostini:2018tnm},
muon-positron ($\mu^-$-$e^+$) conversion processes~\cite{Kaulard:1998rb,Bartoszek:2014mya,Adamov:2018vin}, rare meson decays~\cite{Miyazaki:2012mx,CortinaGil:2019dnd,LHCb,BABAR:2012aa}, and so on. 
Currently, the $0\nu\beta\beta$ experiments give the stringent upper bound on the absolute value of $(m_{\nu})_{ee}$, the $(e,\ e)$ element of the neutrino mass matrix~\cite{KamLAND-Zen:2016pfg},
which is so-called the effective neutrino mass. 
In addition, new further $0\nu\beta\beta$ experiments being planned,
and some of them will reach the lower limit of 
$|(m_{\nu})_{ee}|$ for the scenario of the inverted hierarchy~\cite{Shirai:2018ycl}.
The $\mu^-$-$e^+$ conversion processes were searched at the SINDRUM-II experiment~\cite{Kaulard:1998rb}.
Some next generation experiments are going to be performed~\cite{Bartoszek:2014mya,Adamov:2018vin}.
In addition, there are some experiments searching the LNV decays of charged mesons or $\tau$ lepton~\cite{Miyazaki:2012mx,CortinaGil:2019dnd,LHCb,BABAR:2012aa}.

LNV phenomena can also be tested at future collider experiments. 
In Ref.~\cite{deGouvea:2007qla}, this possibility has been studied in the EFT approach. 
Searching new particles which cause the LNV at hadron collider experiments has also been studied in various ultraviolet (UV) complete models, such as the Type I seesaw model~\cite{Atre:2009rg}, the Type II seesaw model~\cite{Akeroyd:2005gt, Perez:2008ha,Akeroyd:2007zv}, the Left-Right symmetric model~\cite{Huitu:1996su, Das:2012ii} , and so on~\cite{Cai:2017mow, Deppisch:2015qwa}. 
The signature of the Majorana nature at $e^+e^-$ or $e^-e^-$ collider experiments has been studied in Refs.~\cite{Atwood:2007zza, Grimus:2009sq, Aoki:2010tf, Banerjee:2015gca}.
Experimental searches for the LNV at the Large Hadron Collider (LHC) are in Refs.~\cite{Aaboud:2018spl, Sirunyan:2018xiv}.
In~2018, the same-sign W boson fusion process was observed at the LHC~\cite{cite:W_pair_fusion}.
We expect that, in the near future, we can test the LNV signal from the same-sign lepton pair production via the same-sign W boson fusion processes $pp\to W^+W^+jj \to \ell^+\ell^+j j$.

In this paper,
we study impacts of dimension-five LNV operators associated with two same-sign weak bosons, $\ell^\pm \ell^{\prime \pm} W^\mp W^\mp$, on current and future experiments for neutrino oscillation, LNV rare processes and  high energy collider experiments.
These operators can contain important information on the origin of tiny neutrino masses, which is independent of that from the Weinberg operator. 
We examine constraints on the coefficients of the LNV operators by the neutrino oscillation data.
Upper bounds on the coefficients are also investigated using the data for LNV processes such as neutrinoless double beta decays and $\mu^-$-$e^+$ conversion. 
These operators can be directly tested by the lepton number violating processes via the same-sign W boson fusion process at high energy hadron colliders, like the LHC.
It is found that these operators can be considerably probed by these current and future experiments.

This paper is organized as follows.
In Sec.~\ref{sec:operators}, we define the dimension-five LNV operators, $\ell^\pm \ell^{\prime \pm} W^\mp W^\mp$, and discuss the relation to the operators symmetric under the electroweak gauge symmetry.
In Sec.~\ref{sec:Neutrino masses}, we consider neutrino masses which are generated by the $\ell^\pm \ell^{\prime \pm} W^\mp W^\mp$ operators.
They are generated at loop level and have UV divergences from loop integrals. 
We show that these divergences can be renormalized by using higher-dimensional counter terms at the loop level we calculate, 
and we can use the data of neutrino mass matrix as the input parameters of the renormalization procedure. 
In Sec.~\ref{sec:Constraints from low energy experiments}, we derive tree-level constraints for $\ell^\pm \ell^{\prime \pm} W^\mp W^\mp$ from neutrinoless double beta decays and muon positron conversion.
In Sec.~\ref{sec:Collider signatures}, we investigate the LNV signal via the $\ell^\pm \ell^{\prime \pm} W^\mp W^\mp$ operators at the LHC.
Conclusions are given in Sec.~\ref{sec:Conclusion}.
In Appendix~\ref{appendix:dim-7_model}, we show two renormalizable models
which realize the \llww operators with left-handed charged leptons via gauge invariant dimension-seven LNV operators.
In Appendix~\ref{appendix:Renormalization of two point functions of neutrinos},
detailed calculations for the renormalization of two-point functions of neutrinos are shown.

\section{Gauge symmetric operators which yield $\ell^\pm \ell^{\prime \pm} W^\mp W^\mp$}
\label{sec:operators}
We here introduce the dimension-five $\ell^\pm \ell^{\prime \pm} W^\mp W^\mp$ operators, where $\ell(\ell^\prime)$ is a charged lepton $e$, $\mu$ or $\tau$, and $W^\pm$ are the weak bosons. 
Such operators are, in general, represented by the following form\footnote{We do not consider the operators which include derivatives because they have higher dimensions than five.};
\bal
\overline{ \ell^c }\, \Gamma^{\mu \nu} \ell^\prime\, W_\mu^+ W^+_\nu,
\hspace{10pt}
\overline{ \ell^\prime }\, \Gamma^{\mu \nu} \ell^c\, W_\mu^- W^-_\nu,
\eal
where $\Gamma^{\mu\nu}$ is a $4\times4$ matrix which is the product of gamma matrices.
We can classify $\Gamma^{\mu\nu}$ into four forms.
\bal
\Gamma^{\mu\nu} = \left\{
\begin{array}{l}
g^{\mu\nu} P_X, \\
\left[\gamma^\mu, \gamma^\nu\right]P_X,\\
\end{array}
\hspace{10pt} X = L\ \text{or}\ R, 
\right.
\eal
where $P_X$ is the chirality projection operator and $X$ is the chirality of charged leptons.
The operators with the anti-symmetric tensor $\Gamma^{\mu\nu} = \left[\gamma^\mu, \gamma^\nu\right]P_X$ equal zero, because $W^+_\mu W^+_\nu$ is the symmetric for the exchange $\mu \leftrightarrow \nu$.
Therefore, the $\ell^\pm \ell^{\prime \pm} W^\mp W^\mp$ operators are expressed as
\bal
\mathcal{L}_{\sf eff}^{\ell \ell WW} = \sum_{\ell, \ell^\prime} \sum_{X}
\frac{ C^{ X }_{ \ell \ell^\prime } }{ \Lambda } \, \overline{ \ell{}^c } \, P_X \, \ell^\prime \, W_\mu^+ \, W^{+ \mu}
+ \mathrm{ h.c. },
\label{eq:llWWoperators}
\eal
where $C_{\ell \ell^\prime}^X$ are dimensionless coupling constants, and $\Lambda$ is a dimensionful parameter.
The $\mathrm{SU(2)_L \times U(1)_Y}$ gauge invariant origins of theses operators in Eq.~(\ref{eq:llWWoperators}) depend on the chirality $X$ as discussed in order below.

The gauge invariant origin of the $\ell^\pm \ell^{\prime \pm} W^\mp W^\mp$ operators for left-handed charged leptons, $X=L$, is the dimension-seven operators~\cite{Lehman:2014jma,cite:Wudka},
\bal
\frac{ C^{ ( 7 ) }_{\ell \ell^\prime} }{ \Lambda_\mathrm{LNV}^3 }
\left( \, \overline{ \tilde{ L }_{ \ell } } D_\mu L_{ \ell^\prime } \right)
\left( \tilde{ \phi}^\dagger D^\mu \phi \right)
+ \mathrm{ h.c. },
\label{eq:dim7LNV}
\eal
where $\phi$ is the Higgs doublet field in the SM and $\tilde{\phi}$ is its $\mathrm{SU(2)_L}$ conjugation, $L_\ell$ are lepton doublet fields and $\widetilde{L}_\ell$ are their $\mathrm{SU(2)_L}$ conjugations, $C_{\ell \ell^\prime}^{ ( 7 ) }$ are dimensionless coefficients, and $\Lambda_\mathrm{LNV}$ is the scale of lepton number violation. After the electroweak symmetry breaking, the neutral component $\phi^0$ of the Higgs field obtains the vacuum expectation value 
$\left< \phi^0 \right> = v/\sqrt{2}$ with $v = 246\ \mathrm{GeV}$, and the following dimension-five operators are generated;
\bal
- \frac{ i e }{ 2\sqrt{2} s_\mathrm{w} } 
\frac{ v^2 }{ \Lambda_\mathrm{LNV}^2 }
\frac{ C^{ ( 7 ) }_{\ell \ell^\prime} }{ \Lambda_\mathrm{LNV} }
&
\left[ \relax 
	(\, 
	  	\overline{ \ell_L{}^c } \, \partial_\mu \, \nu_{\ell^\prime,L}
	 	 - \overline{ \nu_{\ell,L}^c }\, \partial_\mu \, \ell^\prime_L 
	\, ) W^{+\mu}
	- \frac{ i e }{ \sqrt{ 2 } s_\mathrm{w} } \, \overline{ \ell_L{}^c }\, \ell_L^\prime\, W_{\mu}^+\, W^{+ \mu}
\right.
\nonumber \\
&
\left.
	+ \frac{ i\, e }{ \sqrt{ 2 } s_\mathrm{w} }\, \overline{ \nu_{\ell,L}^c}\, \nu_L^{ \ell^\prime }\, W_{\mu}^-\, W^{+ \mu}
	- \frac{ i\, e }{ 2 s_\mathrm{w} c_\mathrm{ w } }\, \overline{ \ell_L{}^c }\, \nu_{\ell^\prime,L}\, Z_\mu W^{ + \mu }
\right.
\nonumber \\
&
\left.
	-  i\, e \cot 2 \theta_\mathrm{w}\,  \overline{ \nu_{\ell,L}^c }\, \ell_L^\prime\, Z_\mu W^{ + \mu }
	-i \, e \, \overline{ \nu_{\ell,L}^c }\, \ell_L^\prime \, A_\mu W^{ + \mu }
\right] \relax
+ \mathrm{ h. c. },
\label{eq:dim7LNV_expanded}
\eal
where $e$ is the gauge coupling constant of the electromagnetic force, $s_\mathrm{ w } = \sin \theta_\mathrm{w}$, $ c_\mathrm{w} = \cos \theta_\mathrm{ w }$ with $\theta_\mathrm{ w }$ being the Weinberg angle.
The second term in the first row of Eq.~(\ref{eq:dim7LNV_expanded}) corresponds to the operator in Eq.~(\ref{eq:llWWoperators}).
The coupling constants defined in Eq.~(\ref{eq:llWWoperators}) are given by
\bal
\frac{ C_{\ell \ell^\prime}^L }{ \Lambda } 
=
- \frac{ e^2 }{ 4 s_\mathrm{ w }^2 } \frac{ v^2 }{ \Lambda_\mathrm{LNV}^2 }
\frac{ 1 }{ \Lambda_\mathrm{LNV} }
\left(
	 \frac{ C_{\ell \ell^\prime }^{ ( 7 ) } + C_{ \ell^\prime \ell }^{ ( 7 ) } }{ 2 }
\right).
\label{eq:def_CL}
\eal
We note that the original coupling constants $C_{ \ell \ell^\prime }^{ ( 7 ) }$ are not symmetric for flavor indices generally while $C_{\ell \ell^\prime }^L$ are symmetric.
In Appendix~\ref{appendix:dim-7_model}, we show concrete models where the dimension-seven operators in Eq.~(\ref{eq:dim7LNV}) are yielded at one-loop level. 

Next, we consider the gauge invariant origin of the $\ell^\pm \ell^{\prime\pm} W^\mp W^\mp$ operators for right-handed charged leptons, $X=R$.
Contrary to the case of left-handed charged leptons, 
they are generated from the dimension-nine gauge invariant LNV operators~\cite{Gustafsson:2014vpa,cite:Wudka},
\bal
\frac{ C_{ \ell \ell^\prime }^{ ( 9 ) } }{ \Lambda_\mathrm{ LNV }^5 }\,
\overline{ \ell_R{}^c }\, \ell_R^\prime
\left(\, 
	\tilde{ \phi }^\dagger D_\mu \phi\,
\right)^2,
\label{eq:dim9LNV}
\eal
where $\ell_R$ are a right-handed charged lepton,
$C_{\ell \ell^\prime }^{ ( 9 ) }$ are the dimensionless coupling constants. 
After the electroweak symmetry breaking, the dimension-five operators
\bal
- \frac{ e^2 }{ 8 s_\mathrm{ w }^2 }
\frac{ v^4 }{ \Lambda_\mathrm{ LNV}^4 }
\frac{ C_{ \ell \ell^\prime }^{ ( 9 ) } }{ \Lambda_\mathrm{ LNV } }\,
\overline{ \ell_R{}^c }\, \ell_R^\prime\, W_{\mu}^+ W^{ + \mu },
\eal
are generated. 
Therefore, the coupling constants $C_{\ell \ell^\prime}^R$ can be expressed by the parameters of gauge invariant effective LNV operators as
\bal
\frac{ C_{ \ell \ell^\prime }^R }{ \Lambda }
= 
- \frac{ e^2 }{ 8 s_\mathrm{ w }^2 }
\frac{ v^4 }{ \Lambda_\mathrm{ LNV}^4 }
\frac{ C_{ \ell \ell^\prime }^{ ( 9 ) } }{ \Lambda_\mathrm{ LNV } }.
\label{eq:def_CR}
\eal
Notice that new coupling constants $C_{ \ell \ell^\prime }^R$ are symmetric for flavor indices because $C_{ \ell \ell^\prime }^{ ( 9 ) }$ are symmetric.
In Refs.~\cite{delAguila:2011gr,Gustafsson:2012vj, Gustafsson:2014vpa,cite:Wudka}, the models where the dimension-nine operators in Eq.~(\ref{eq:dim9LNV}) are yielded at tree or one-loop level are investigated.

\section{Neutrino masses}
\label{sec:Neutrino masses}

In addition to the Weinberg operator, the LNV operators in Eqs.~(\ref{eq:dim7LNV}) and~(\ref{eq:dim9LNV}) can contribute to Majorana masses of neutrinos at loop levels. 
The coefficients of these operators are constrained  by
the current data for the neutrino mass matrix which is given by neutrino oscillation experiments and observation of cosmic microwave background.

We begin with summarizing the observed results for the neutrino mass matrix.
The Majorana-type mass matrix~$m_\nu$ is diagonalized by using a unitary matrix, so-called the Pontecorvo-Maki-Nakagawa-Sakata (PMNS) matrix~$U$~\cite{cite:Pontecorvo,Maki:1962mu},
\bal
m_\nu = U 
\left(
\begin{array}{ccc}
m_1 & 0 & 0 \\
0 & m_2 & 0 \\
0 & 0 & m_3 \\
\end{array}
\right)
U^\mathrm{T},
\eal
where $m_i\  (i =1,2,3)$ is a mass eigenvalue of $m_\nu$.
The PMNS matrix can be parametrized as follows;
\bal
U = 
 \begin{pmatrix}
  1 & 0 & 0 \\
  0 & c_{23} & s_{23} \\
  0 & -s_{23} & c_{23} \\
 \end{pmatrix}
 \begin{pmatrix}
  c_{13} & 0 & s_{13}e^{-i\delta} \\
  0 & 1 & 0 \\
  -s_{13}e^{i\delta} & 0 & c_{13} \\
 \end{pmatrix}
 \begin{pmatrix}
  c_{12} & s_{12} & 0 \\
  -s_{12} & c_{12} & 0 \\
  0 & 0 & 1 \\
\end{pmatrix}
\begin{pmatrix}
1 & 0 & 0 \\
0 & e^{i\alpha_1} & 0 \\
0 & 0 & e^{i \alpha_2} \\
\end{pmatrix},
\label{eq:PMNS_parametrization}
\eal
where $c_{ij}$ and $s_{ij}$ are $\cos \theta_{ij}$ and $\sin \theta_{ij}$ with  
$\theta_{12},\theta_{13}$ and $\theta_{23}$ being mixing angles, and
$\delta, \alpha_1$ and $\alpha_2$ are CP violating phases.
We note that $\alpha_1$ and $\alpha_2$ can only exist in the case that neutrinos are Majorana fermions. 
We here list the values of each parameter with $1\sigma$ errors which are observed by neutrino oscillation experiments~\cite{pdg}; 
\bal
&\Delta m_{21}^2 = m_2^2 - m_1^2 = (7.53 \pm 0.18) \times 10^{-5}\ \mathrm{eV^2},
\nonumber \\
&\sin^2 \theta_{12} = 0.307 \pm 0.013 ,
\nonumber \\
&\Delta m_{32}^2 = m_3^2 - m_2^2 = (2.444 \pm 0.034)\ \times 10^{-3}\ \mathrm{eV^2}
\ \text{(for NH)},
\nonumber\\
&\Delta m_{32}^2 =  (-2.55 \pm 0.04) \times 10^{-3}\ \mathrm{eV^2}
\ \text{(for IH)},
\nonumber \\
&\sin^2 \theta_{23} = 0.512^{+ 0.019}_{-0.022}\ \text{(for NH)},
\nonumber \\
&\sin^2 \theta_{23} = 0.536^{+0.023}_{-0.028}\ \text{(for IH)},
\nonumber \\
&\sin^2 \theta_{13} = (2.18 \pm 0.07)\times 10^{-2},
\label{eq:Constraint_from_neutrino_oscillation}
\eal
where NH and IH are abbreviations of normal hierarchy and inverted hierarchy, respectively.
The CP violating phases $\alpha_1$ and $\alpha_2$ cannot be observed at neutrino oscillation experiments, and we do not have any information of them currently.
The rest CP violating phase $\delta$ can be observed at neutrino oscillation experiments, and latest data at T2K have already ruled out the CP conserving cases ($\delta = 0$ or $\pi$) at $99.73\%$ C.L.~\cite{Abe:2019vii}.
We can measure the difference between quadratics of their mass eigenvalues by neutrino oscillation experiments, however, the pattern of the hierarchy, whether the normal hierarchy where $m_1 < m_2 < m_3$ or the inverted hierarchy where $m_3 < m_1 < m_2$, is still unknown.
Although absolute values for each mass eigenvalue cannot be determined  by neutrino oscillation experiments, 
the upper bound for the summation of the mass eigenvalues can be obtained from observation of the cosmic microwave background. 
From the latest result by the Planck collaboration~\cite{Aghanim:2018eyx}, the following constraint is given;
\bal
m_1 + m_2 + m_3 < 0.12\ \mathrm{eV} \hspace{10pt} (95\%\, \mathrm{C.L.}).
\label{eq:Constraint_from_CMB}
\eal
Information on absolute values of neutrino masses can also be given by $0\nu\beta\beta$ experiments which can constrain the $(e,e)$ component
$|(m_\nu)_{ee}|$ of the effective neutrino mass,
which is discussed in more details in Section~\ref{sec:Constraints from low energy experiments}.

\begin{figure}[h]
\begin{center}
\includegraphics[width=150mm]{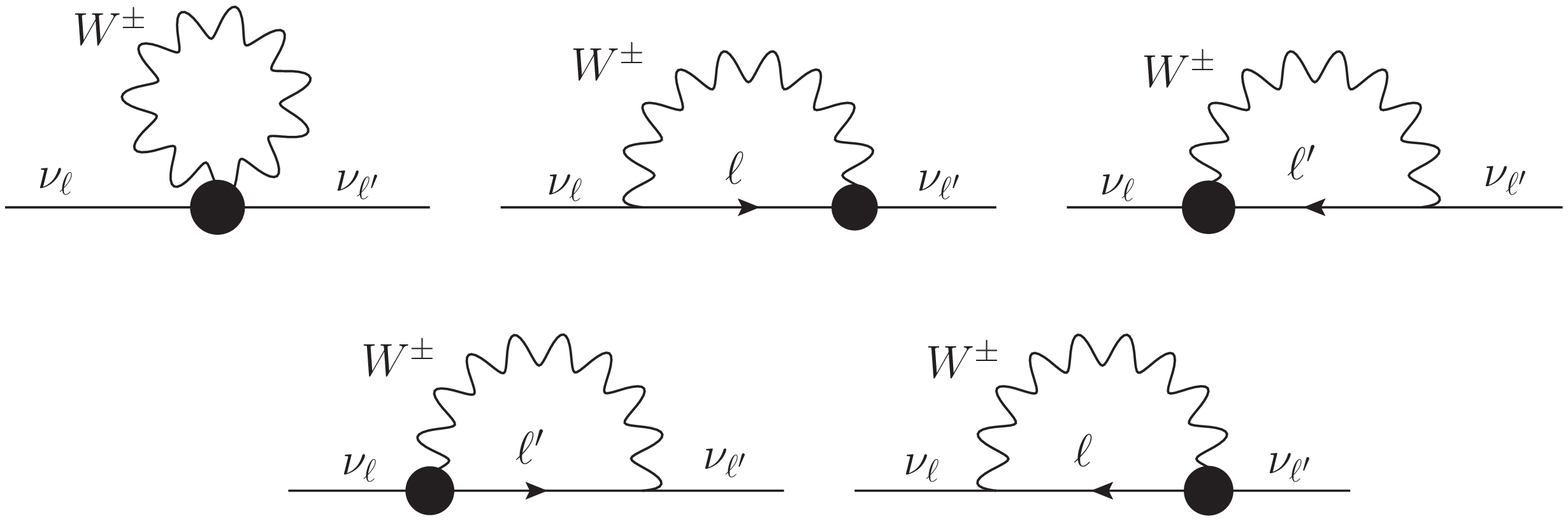}
\caption{Feynman diagrams for neutrino masses which are generated by the dimension-seven LNV operators.}
\label{fig:MassLH}
\vspace{50pt}
\includegraphics[width=130mm]{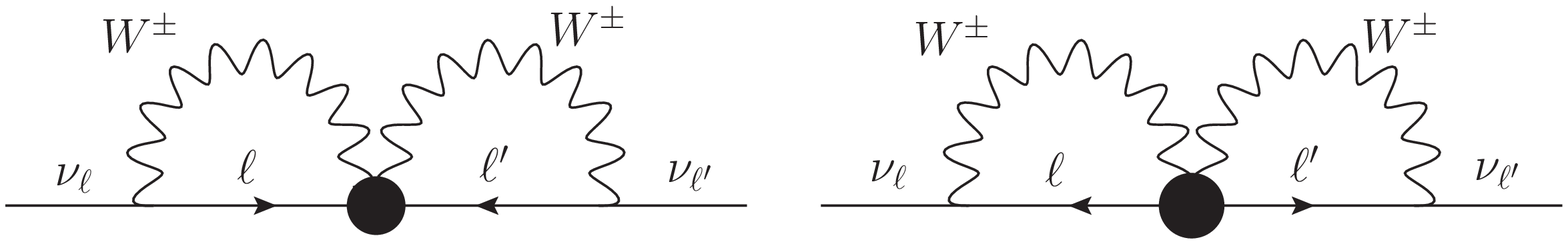}
\caption{Feynman diagrams for neutrino masses which are generated by the dimension-nine LNV operators.}
\label{fig:MassRH}
\end{center}
\end{figure}

We now consider the constraint on the coefficients of the \llww operators.
At tree level, neutrino masses would be generated via the Weinberg operator,
\bal
\frac{ C_{\ell \ell^\prime}^{(5)} }{ \Lambda_5 }
\Bigl(
	\overline{ \tilde{L}_\ell } \phi
\Bigr)
\Bigl(
	\tilde{\phi}^\dagger L_{\ell^\prime}
\Bigr) + \mathrm{h.c.},
\label{eq:Weinberg_operator}
\eal
where $\Lambda_5$ is the scale of physics where the Weinberg operator is generated.
In general, $\Lambda_5$ can be different from $\Lambda_\mathrm{LNV}$,
depending on the scenario of creating tiny neutrino masses.
At loop level, operators in Eqs.~(\ref{eq:dim7LNV}) and ~(\ref{eq:dim9LNV}) can contribute to neutrino masses.
We show Feynman diagrams for neutrino masses in Figs.~\ref{fig:MassLH} and~\ref{fig:MassRH}.
These loop diagrams have UV divergences which are caused by using the higher-dimensional operators.
We use the renormalization procedure to eliminate these divergences. 
Although theories which include the higher-dimensional operators are not renormalizable at all orders of perturbation, 
the divergence at one-loop (two-loop) level of Feynman diagrams in Fig.~\ref{fig:MassLH} (Fig.~\ref{fig:MassRH}) can be renormalized by using the higher-dimensional counter terms. 
Then, the observed data of the neutrino mass matrix are used to impose the renormalization conditions.  
With this renormalization procedure, we can handle at least one- or two-loop terms in the loop expansion consistently. 
In the following, we show the outline and the result of this procedure. 
The details of the renormalization calculation are shown in Appendix~\ref{appendix:Renormalization of two point functions of neutrinos}. 

First, we consider the renormalization of two-point functions of neutrinos which are generated via the dimension-seven operators given in Eq.~(\ref{eq:dim7LNV}).
Diagrams in Fig.~\ref{fig:MassLH} have quadratic divergences and logarithmic divergences proportional to the squared momentum of external neutrinos.
We can eliminate the former one (the quadratic divergence) by using the counter term from the Weinberg operator.
In order to eliminate the latter one (the logarithmic one), we use the new counter term from the dimension-seven operators~\cite{cite:Wudka},
\bal
F^{(7)}_{\ell \ell^\prime} 
 \left( \overline{ L }_\ell  \tilde{ \phi } \right)
\partial_\mu \left( \phi^\dagger D^\mu \widetilde{L}_{\ell^\prime} \right) + \mathrm{h.c.},
\label{eq:dim7counter}
\eal
where $F^{(7)}_{\ell \ell^\prime}$ are the coupling constants whose mass dimension is $-3$.
We use the data for the neutrino mass matrix to impose the on-shell renormalization condition to the two-point function of neutrinos.
After this renormalization procedure, we obtain the renormalized amputated two-point function of neutrinos in the mass eigenstate basis as follows;
\bal
i \Sigma_{ab}(\cancel{p})
=& \left<0| \mathcal{T} \nu_a \overline{\nu_b} |0\right>_\mathrm{amp}
\nonumber \\
=&\  i \Sigma_{ab}^L(\cancel{p}) P_L
	 + i\Bigl( \Sigma_{ab}^L(\cancel{p}) \Bigr)^\ast P_R,
\label{eq:def_SigmaL}
\\
\Sigma_{ab}^L(\cancel{p})
\simeq &
\ - \frac{1}{ 16\pi^2 }\, U_{\ell a} \frac{ C_{ \ell \ell^\prime }^L }{ \Lambda }\, U_{\ell^\prime b}
\, f\left(\frac{p^2}{m_W^2}\right),
\label{eq:prediction_SigmaL}
\\
f(x) = & \ \frac{ 1 }{ 36 x^2 } \Bigl( x(6+57x-97x^2) + 6(1-x)^2 (11x+1) \ln(1-x) \Bigr),
\label{eq:f(x)}
\eal
where we use Eq.~(\ref{eq:def_CL}), and $m_W$ is the mass of the weak bosons $W^\pm$, $p_\mu$ is the momentum of the external neutrino, and neutrino fields in the mass eigenstate basis are defined as
\bal
\label{eq:Majorana_neutrino1}
&\nu_{a} = \nu_{a,L} + \nu_{a,L}^c, \\
\label{eq:Majorana_neutrino2}
&\nu_{a,L} = U_{\ell a}^\ast \nu_{\ell,L}.
\eal
In Eq.~(\ref{eq:prediction_SigmaL}), we only show the leading term,
neglecting terms proportional to the masses of charged leptons.
Details of the calculation are shown in Appendix~\ref{appendix:Renormalization of two point functions of neutrinos}.
Then, neutrino mass eigenvalues and mixing angles are input parameters,
and the coefficients $C_{\ell \ell^\prime}^L/\Lambda$ are not constrained from the data of neutrino oscillation.

\begin{figure}[b]
\begin{center}
\includegraphics[width=100mm]{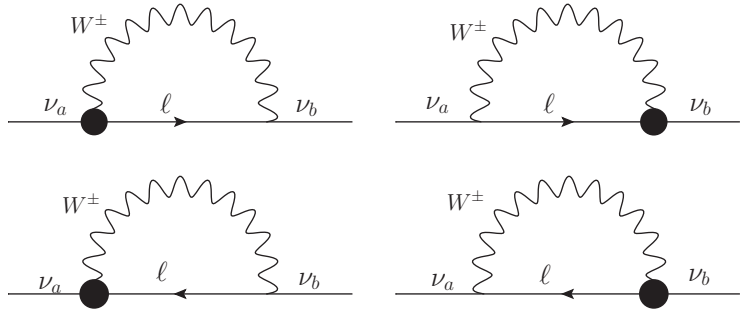}
\caption{Two-point functions which are generated by the LNV operators in Eq.~(\ref{eq:dim7counter2}).}
\label{fig:MassDiagram_for_Renormalization}
\end{center}
\end{figure}

Next, we consider the renormalization of two-point functions of neutrinos which are generated via dimension-nine operators given in Eq.~(\ref{eq:dim9LNV}).
In order to eliminate all divergences which appear in the Feynman diagrams in Fig.~\ref{fig:MassRH}, 
we introduce new dimension-seven operators, 
\bal
F^{\, \prime (7)}_{\ell \ell^\prime} ( \phi^\dagger D^\mu \widetilde{ \phi } ) ( \phi^\dagger \overline{\ell_R } \gamma_\mu \widetilde{ L }_{ \ell^\prime } ) + \mathrm{h.c.},
\label{eq:dim7counter2}
\eal
where $F_{\ell \ell^\prime}^{\, \prime (7)}$ are the coupling constants whose mass dimension is $-3$.
We use three LNV operators in Eqs.~(\ref{eq:dim7LNV}),~(\ref{eq:Weinberg_operator}) and~(\ref{eq:dim7counter2}).
At one-loop level, Majorana masses of neutrinos are generated via the dimension-seven operators in Eq.~(\ref{eq:dim7counter2}). 
Feynman diagrams are shown in Fig.~\ref{fig:MassDiagram_for_Renormalization}.
These diagrams have logarithmic divergences.
These divergences can be renormalized by using 
$\mathcal{O}(\hbar)$ counter terms from the Weinberg operator and the on-shell renormalization conditions in Appendix~\ref{appendix:Renormalization of two point functions of neutrinos}.
At two-loop level, the dimension-nine operators in Eq.~(\ref{eq:dim9LNV})
generate the Majorana masses of neutrinos via the Feynman diagrams 
in Fig.~\ref{fig:MassRH}.
These diagrams have two kinds of divergences; i.e., logarithmic divergences and squared logarithmic divergences.
The squared logarithmic divergences can be eliminated by using 
$\mathcal{O}(\hbar^2)$ counter terms from the Weinberg operator.
The logarithmic divergences are proportional to a function of the momentum of the external neutrino.
In order to eliminate these divergences, we use $\mathcal{O}(\hbar^2)$ 
counter terms from the operators in Eq.~(\ref{eq:dim7counter2}).
After this renormalization procedure with renormalization conditions in Appendix~\ref{appendix:Renormalization of two point functions of neutrinos},
we obtain the renormalized amputated two-point functions of neutrinos in the mass eigenstate basis as
\bal
\label{eq:predictionSigmaLRH}
\Sigma^L_{ab}(\cancel{p})
\simeq &
- \frac{ i v^3 }{ 32 \pi^2 } \left( \frac{ e }{ s_\mathrm{w} } \right)^2
 \Bigl\{
 	(U^\mathrm{T} F^{\prime(7)})_{b\ell} \, m_\ell \, U_{\ell a} 
	+ (U^\mathrm{T} F^{\prime(7)})_{a \ell} \, m_\ell \, U_{\ell b} 
\Bigr\}
\, g\left( \frac{ p^2 }{ m_W^2 } \right)
\nonumber \\
&- \frac{ 1 }{ 128\pi^4 } \left( \frac{ e }{ s_\mathrm{w} } \right)^2
\, \frac{ C_{\ell \ell^\prime}^R }{\Lambda}\, U_{\ell a}\, U_{\ell^\prime b}\, 
m_\ell\, m_{\ell^\prime}\, \biggl\{ g\left( \frac{ p^2 }{ m_W^2 } \right) \biggr\}^2,
\\
g\left( x \right) = & \ 1 + \frac{ (1-x) \ln( 1-x) }{ x },
\eal
where $\Sigma_{ab}^L(\cancel{p})$ are defined in Eq.~(\ref{eq:def_SigmaL}),
and we use Eq.~(\ref{eq:def_CR}).
In Eq.~(\ref{eq:predictionSigmaLRH}), we only show the leading term, neglecting terms proportional to cubic or higher order terms of charged lepton masses. 
Detail of the calculation are shown in Appendix~\ref{appendix:Renormalization of two point functions of neutrinos}.
As in the case for $C_{\ell \ell^\prime}^L/\Lambda$, the coefficients $C_{\ell \ell^\prime}^R/\Lambda$ are not constrained from the data of neutrino oscillation.

In the above renormalization procedure, 
we do not have constraints on \llww 
operators from the observed data of neutrino oscillation as a result. 
However, it does not mean that we do not have any prediction for new physics. 
When we consider the LNV processes at one-loop or two-loop level, 
the renormalized LNV operators can give some prediction
under constraints from the neutrino oscillation data. 
In the following sections, we investigate the LNV processes only at tree level. 
Therefore, the constrains from the neutrino oscillation data are not important in the discussions below.

\section{Constraints from low energy experiments}
\label{sec:Constraints from low energy experiments}

In this section, we discuss current constraints on the \llww operators from low energy experiments; i.e., neutrinoless double beta decays (\,$0\nu\beta\beta$\,) and muon-positron ($\mu^-$-$e^+$) conversion processes. 
The constraint on LNV higher-dimensional operators from $0\nu\beta\beta$ ($\mu^-$-$e^+$ conversion) are studied in Refs.~\cite{cite:Wudka, Cirigliano:2017djv} (Refs.~\cite{Berryman:2016slh, deGouvea:2019xzm, Geib:2016atx}).
The neutrinoless double beta decay via the $e^-e^-W^+W^+$ operator in the UV complete models 
is discussed in Refs.~\cite{Gustafsson:2014vpa, delAguila:2011gr}.

\subsection{ Neutrinoless double beta decay (\,$ 0 \nu \beta \beta $\,) }
We consider the constraint from the $0\nu\beta\beta$\ experiments.
Currently, KamLAND-Zen experiment provides the most stringent limit on the half-life of the process at $90\%$ C.L.~\cite{KamLAND-Zen:2016pfg},
\bal
T_{1/2} >  1.07 \times 10^{26}\ \mathrm{years}.
\label{eq:halftime_KamLAND}
\eal
If we assume that the process occurs via Majorana masses of neutrinos, this bound is translated to the upper limit on the absolute value of the $(e,e)$ element of the effective neutrino mass matrix at $90\%$ C.L.~\cite{KamLAND-Zen:2016pfg},
\bal
|(m_\nu)_{ee}| < ( 61 - 165 )\ \mathrm{meV}.
\label{eq:eff_numass_KamLAND}
\eal
We can then estimate the upper bound on the parton-level amplitude for $dd\to uue^-e^-$,
\bal
\left| \mathcal{M}_{m_\nu}^{0\nu\beta\beta} \right| \simeq & \ \frac{ G_F^2  }{  p_\mathrm{eff}^2 }\, |(m_\nu)_{ee}|,
\label{eq:parton_level_dim5}
\eal 
where $G_F (\simeq 1.17 \times 10^{-5}\;  \mathrm{GeV^{-2}})$ is the Fermi constant and $p_\mathrm{eff} (\sim 100\; \mathrm{MeV})$ is the typical distance scale between nucleons. 
In the following, we extract constraints on $C_{ee}^R/\Lambda$ and $C_{ee}^L/\Lambda$ by comparing
Eq.~(\ref{eq:parton_level_dim5}) to parton-level amplitudes generated by the LNV operators in Eqs.~(\ref{eq:dim7LNV}) and~(\ref{eq:dim9LNV}), respectively.
 
\begin{figure}[h]
 \begin{center}
 \includegraphics[width=60mm]{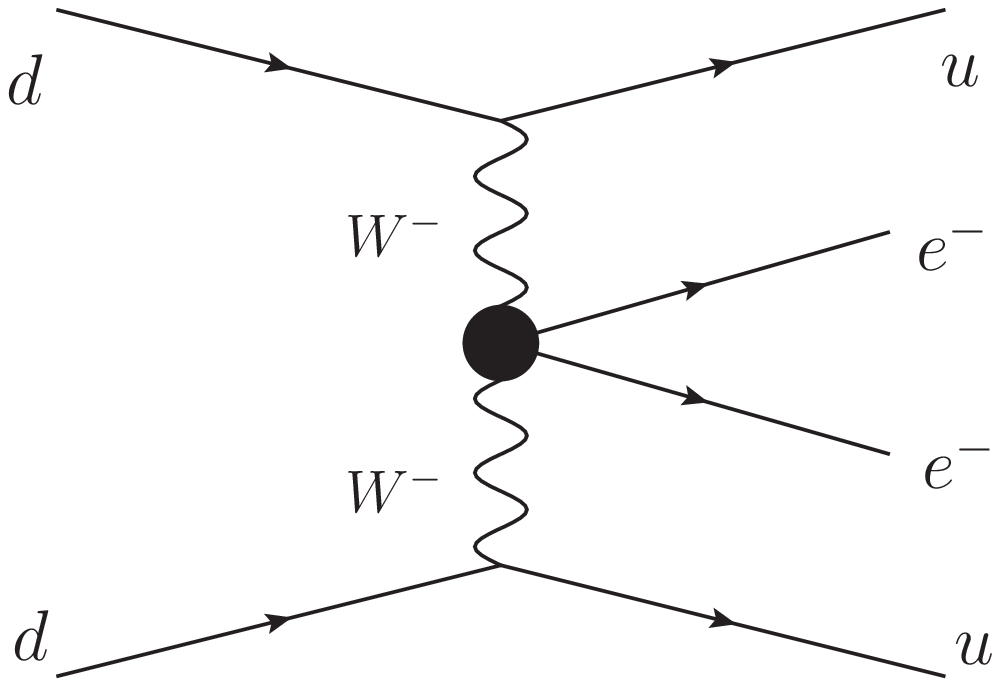}
 \caption{Feynman diagrams for the $0\nu\beta\beta$ decay via the dimension-nine LNV operators.}
 \label{fig:0n2bRH}
 \end{center}
 \end{figure}

First, we consider the constraint on $C_{ee}^R/\Lambda$.
The dimension-nine LNV operators in Eq.~(\ref{eq:dim9LNV}) generate $0\nu\beta\beta$ decays at tree level which are described by the diagram in Fig.~\ref{fig:0n2bRH}.
By using Eq.~(\ref{eq:def_CR}), the parton-level amplitude is given by
 \bal
 \left| \mathcal{M}_R^{0\nu\beta\beta}  \right| \simeq G_F^2 \left| \frac{ C_{ee}^R }{ \Lambda } \right|.
 \label{eq:parton_level_dim7}
 \eal
By comparing this with Eqs.~(\ref{eq:parton_level_dim5}) and~(\ref{eq:parton_level_dim7}), we estimate the upper bound on $C_{ee}^R/\Lambda$ as
\bal
 \left| \frac{ C_{ee}^R }{ \Lambda } \right| \lesssim 10^{-5}\ \mathrm{TeV^{-1}}.
 \label{eq:bound_from_0n2b}
 \eal
We can translate this bound to that on the original coupling constant $C_{ee}^{(9)}/\Lambda_\mathrm{LNV}^5$ by using Eq.~(\ref{eq:def_CR}):
\bal
\left|
	\frac{ C_{ee}^{(9)} }{ \Lambda_\mathrm{LNV}^5 } 
\right| 
\lesssim \, 5.5 \times 10^{-2}\ \mathrm{TeV^{-5}}.
\eal
With the assumption $|C_{ee}^{(9)}|=1$, 
we can obtain the lower bound on the new physics scale $\Lambda_\mathrm{LNV}$ as
\bal
\Lambda_\mathrm{LNV} \gtrsim 1.8\ \mathrm{TeV}.
\eal

\begin{figure}[h]
 \begin{center}
 \includegraphics[width=170mm]{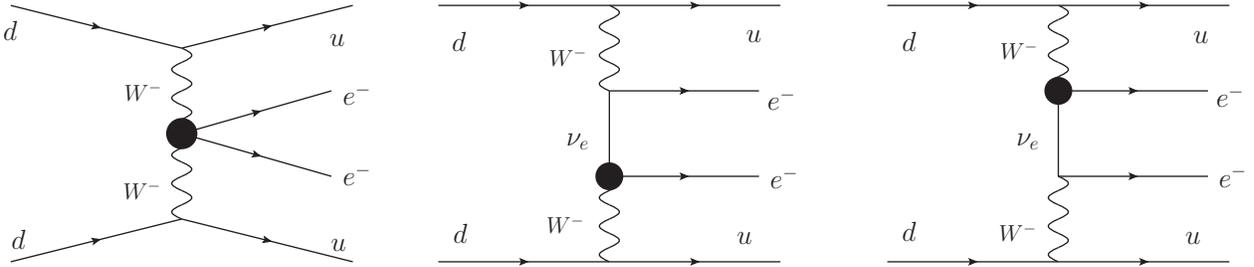}
 \caption{Feynman diagrams for the $0\nu\beta\beta$ decay via the dimension-seven LNV operators.}
 \label{fig:0n2bLH}
 \end{center}
 \end{figure}

Next, we consider the constraint on $C_{ee}^L/\Lambda$.
 The dimension-seven LNV operators in Eq.~(\ref{eq:dim7LNV}) generate $0\nu\beta\beta$ decays at tree level which are described by the Feynman diagrams in Fig.~\ref{fig:0n2bLH}.
In this case, there are additional diagrams which are generated via three-point vertices in the first line of Eq.~(\ref{eq:dim7LNV_expanded}).
They only change the factor of the amplitude.
By using Eq.~(\ref{eq:def_CL}), we can obtain the same upper bound with that on $C_{ee}^R$,
\bal
\left| \frac{ C_{ee}^L }{\Lambda} \right| \lesssim 10^{-5}\ \mathrm{TeV^{-1}}.
 \label{eq:bound_from_0n2bLH}
\eal
We can translate this bound to that on the original coupling constant $C_{ee}^{(7)}/\Lambda_\mathrm{LNV}^3$ by using Eq.~(\ref{eq:def_CL}):
\bal
\left|
	\frac{ C_{ee}^{(7)} }{ \Lambda_\mathrm{LNV}^3 } 
\right| 
\lesssim \, 1.7 \times 10^{-3}\ \mathrm{TeV^{-3}}.
\eal
With the assumption $|C_{ee}^{(7)}|=1$, 
we can obtain the lower bound on the new physics scale $\Lambda_\mathrm{LNV}$ as
\bal
\Lambda_\mathrm{LNV} \gtrsim 8.4\ \mathrm{TeV}.
\eal


\subsection{ Muon to positron (\,$\mu^-$-$e^+$\,) conversion }
We here consider the constraint on the \llww operators from $\mu^-$-$e^+$ conversion experiments.
The current constraint on the ratio of the rate of the $\mu^-$-$e^+$ conversion with that of the muon capture is given by the SINDRUM-II experiment \cite{Kaulard:1998rb} as follows;
\bal
B_{\mu^- e^+} = \frac{ \Gamma( \mu^- + \mathrm{Ti} \to e^+ + \mathrm{Ca} ) }{ \Gamma( \mu^- + \mathrm{Ti} \to \nu_\mu + \mathrm{Sc} ) } < 
\left\{
\begin{array}{l}
1.7 \times 10^{-12}\ (\mathrm{GS},\ 90\% \mathrm{CL}) \\
3.6 \times 10^{-11}\ (\mathrm{GDR},\ 90\%\mathrm{CL}) \\
\end{array}
\right.
. 
\eal
If we assume that the process occurs via Majorana masses of neutrinos, $B_{\mu^- e^+}$ is calculated by~\cite{Domin:2004tk}.
\bal
B_{\mu^- e^+} = ( 1.6 \times 10^{-25} )\,  \frac{ |(m_\nu)_{e\mu}|^2 }{ m_e^2 }.
\eal
From this formula, we can obtain the upper bound on $|(m_\nu)_{e \mu}|$,
\bal
\label{eq:upperbound_for_emu}
|(m_\nu)_{e\mu}| \lesssim 1.6 \times 10^6\ \mathrm{MeV}.
\eal
In order to extract the constraint on the \llww operators from $\mu^-$-$e^+$ conversion data, we use the similar way to the case of $0\nu\beta\beta$.
In the following, we extract constraints on $C_{e\mu}^R/\Lambda$ and $C_{e\mu}^L/\Lambda$ by comparing
the parton level amplitude for $uu\mu^-\to dde^+$,
\bal
\label{eq:mu_e_Majorana}
|\mathcal{M}_{m_\nu}^{\mu^-e^+}| \simeq \frac{G_F^2}{ p_\mathrm{eff}^2 }\, |(m_\nu)_{e\mu}|,
\eal
to those generated by the LNV operators in Eqs.~(\ref{eq:dim7LNV}) and~(\ref{eq:dim9LNV}), respectively.
 \begin{figure}[h]
 \begin{center}
 \includegraphics[width=60mm]{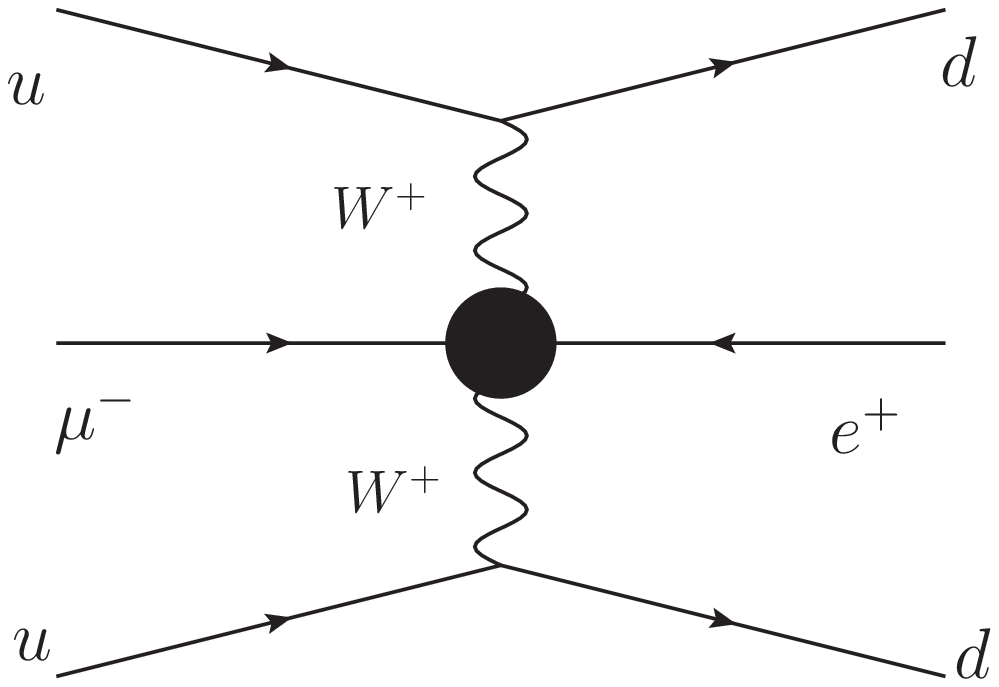}
 \caption{Feynman diagrams for the $\mu^-$-$e^+$ conversion process via the dimension-nine LNV operators.}
 \label{fig:MuERH}
\vspace{50pt}
 \includegraphics[width=170mm]{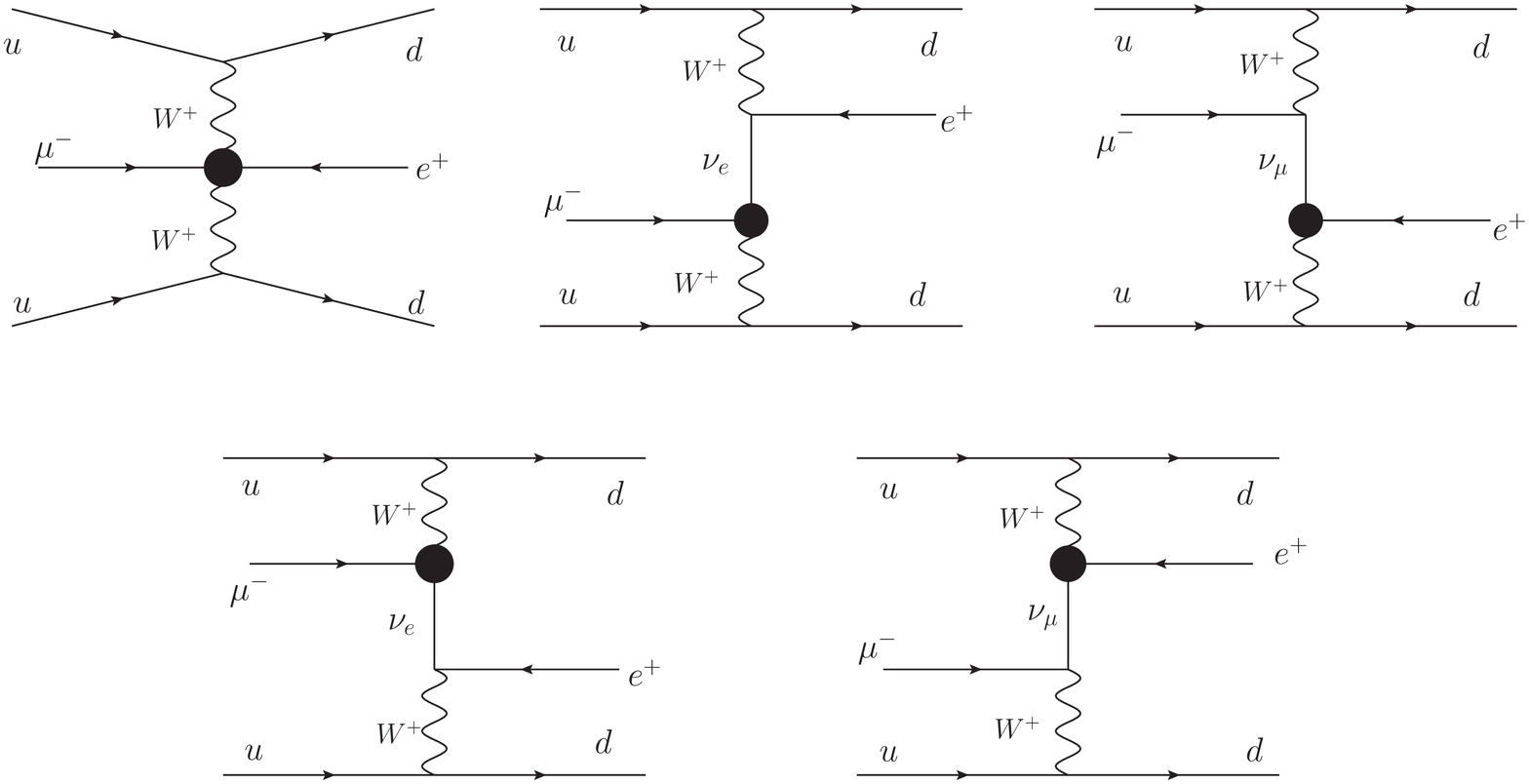}
 \caption{Feynman diagrams for the $\mu^-$-$e^+$ conversion process via the dimension-seven LNV operators.}
 \label{fig:MuELH}
 \end{center}
 \end{figure}

First, we consider the constraint on $C_{ e \mu}^R/\Lambda$.
The dimension-nine operators in Eq.~(\ref{eq:dim9LNV}) generate the $\mu^-$-$e^+$ conversion process at tree level which is described by the diagram in Fig~\ref{fig:MuERH}.
By using Eq.~(\ref{eq:def_CR}), the parton-level amplitude is given by
\bal
|\mathcal{M}_R^{\mu^-e^+}| \simeq G_F^2\, \left|\frac{ C_{e\mu}^R }{ \Lambda } \right|^2.
\eal
By comparing this formula with Eqs.~(\ref{eq:upperbound_for_emu}) and~(\ref{eq:mu_e_Majorana}), we can obtain the constraint on $C_{e\mu}^R/\Lambda$ as
\bal
\left| \frac{ C_{e \mu}^R }{ \Lambda } \right| \, \lesssim \, 1.6 \times 10^8\ \mathrm{ TeV^{-1} }.
\eal
We can translate this bound to that on the original coupling constant $C_{e\mu}^{(9)}/\Lambda_\mathrm{LNV}^5$ by using Eq.~(\ref{eq:def_CR}):
\bal
\left|
	\frac{ C_{e\mu}^{(9)} }{ \Lambda_\mathrm{LNV}^5 } 
\right| 
\lesssim \,  8.8 \times 10^{11}\ \mathrm{TeV^{-5}}.
\eal
With the assumption $|C_{e\mu}^{(9)}|=1$, 
we can obtain the lower bound on the new physics scale $\Lambda_\mathrm{LNV}$ as
\bal
\Lambda_\mathrm{LNV} \gtrsim 4.1\times 10^{-3}\ \mathrm{TeV}.
\eal
This lower bound is too small 
to give a significant constraint on the operator.

Next, we consider the constraint on $C_{e\mu}^L/\Lambda$. In Fig.~\ref{fig:MuELH}, we show the Feynman diagrams which are generated via the dimension-seven LNV operators in Eq.~(\ref{eq:dim7LNV}).
By using Eq.~(\ref{eq:def_CL}), we can get the upper bound as
\bal
\left| \frac{ C_{e \mu}^L }{ \Lambda } \right| \, \lesssim \, 
1.6 \times 10^8\ \mathrm{ TeV^{-1} }.
\eal
We can translate this bound to that on the original coupling constant $(C_{e\mu}^{(7)}+C_{\mu e}^{(7)})/\Lambda_\mathrm{LNV}^5$ by using Eq.~(\ref{eq:def_CL}):
\bal
\left|
	\frac{C_{e\mu}^{(7)} + C_{\mu e}^{(7)} }{2\Lambda_\mathrm{LNV}^3}
\right| 
\lesssim \,2.7 \times 10^{10}\ \mathrm{TeV^{-3}}.
\eal
With the assumption that the symmetric part of $C_{e\mu}^{(7)}$ is $1$, i.e. $(C_{e\mu}^{(7)}+C_{\mu e}^{(7)})/2 = 1$,
we can obtain the lower bound on the new physics scale $\Lambda_\mathrm{LNV}$ as
\bal
\Lambda_\mathrm{LNV} \gtrsim 3.3\times 10^{-4}\ \mathrm{TeV}.
\eal
As in the case of the right-handed operator, 
this lower bound is too small to give a significant constraint on the operator. 

Consequently, both of the right-handed and left-handed \llww operators receive 
almost no constraint from $\mu^-$-$e^+$ conversion. 
In addition, at around the scale $4.1\times 10^{-3}\  \mathrm{TeV}$ or $3.3\times 10^{-4}\, \mathrm{TeV}$,
using the EFT approach might not be a good approximation.

\section{Constraint from high-energy collider experiments}
\label{sec:Collider signatures}

In this section, we investigate LNV processes $pp \rightarrow \ell^+\ell^\prime{}^+j j $ at hadron colliders, and examine the constraints on the LNV coupling constants of the \llww operators. 
The process with the same final state is also studied in UV complete models such as the Type-I seesaw model~\cite{Atre:2009rg} and the Left-Right symmetric model~\cite{Huitu:1996su, Das:2012ii}. 
In the Type-II seesaw model, the same final state can be generated via the decay of the doubly charged scalar. However, the cross section of this process in the Type-II seesaw model is negligibly small because of the tiny neutrino masses, so that other LNV processes are more significant to search the LNV~\cite{Akeroyd:2005gt, Perez:2008ha}. 
\begin{figure}[h]
\begin{center} 
\includegraphics[width=140mm]{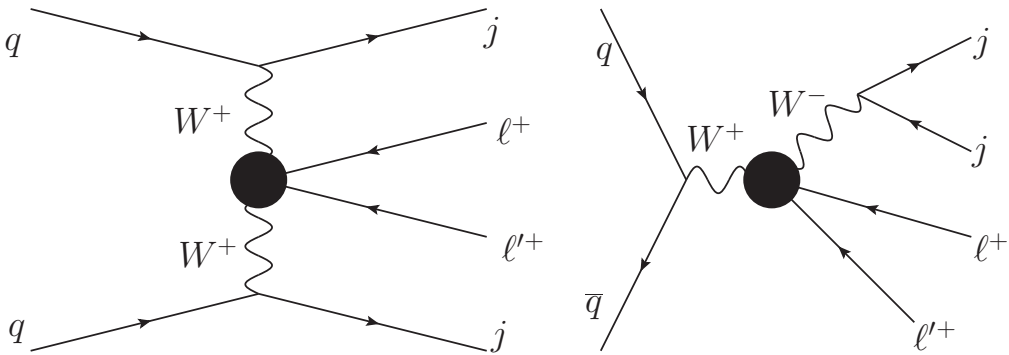}
\caption{Feynman diagrams for the process $pp\rightarrow \ell^+ \ell^\prime{}^+jj $ via the dimension-nine LNV operators.}
\label{fig:lljjRH}
\end{center}
\end{figure}

\subsection{The constraints on $C^R_{\ell \ell^\prime}$}
\begin{figure}[h]
\begin{center}
\includegraphics[width=100mm]{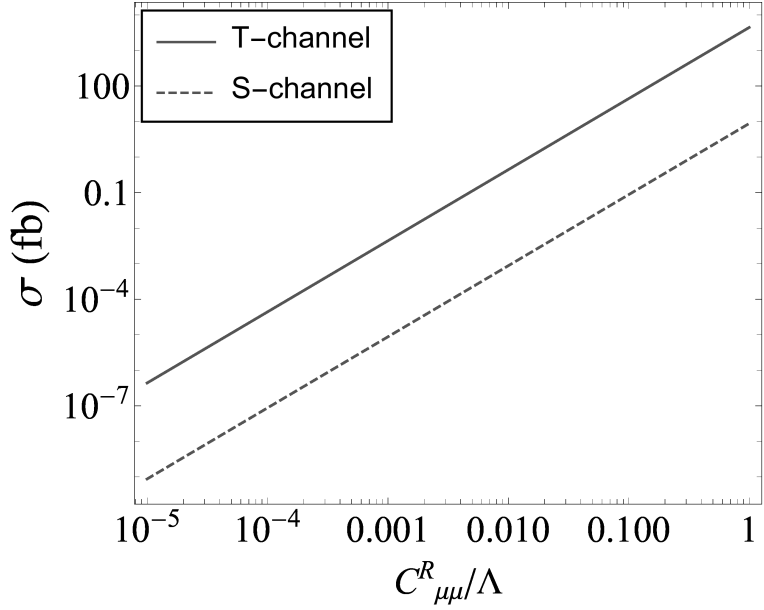}
\caption{ The cross section of the process $pp\to \mu^+\mu^+ j j$. }
\label{fig:Sch_and_Tch}
\end{center}
\end{figure}

We begin with the processes $pp\rightarrow \ell^+\ell^\prime{}^+jj$ with right-handed charged leptons which are generated by the dimension-nine operators in Eq.~(\ref{eq:dim9LNV}).
These processes are represented by the diagrams in Fig.~\ref{fig:lljjRH}.
There are two kinds of processes; t-channel diagrams via W boson fusion processes $qq \rightarrow W^{+(\ast)}W^{+(\ast)}jj\rightarrow \ell^+\ell^\prime{}^+jj $ and s-channel ones $q\overline{q} \to W^{+(\ast)} \to \ell^+\ell^\prime{}^+jj$.
In Fig.~\ref{fig:Sch_and_Tch}, we show the cross section of the process $pp\to \mu^+\mu^+jj$ at $\sqrt{s}=14\ \mathrm{TeV}$ which is calculated by using~FEYNRULES 2.0~\cite{Alloul:2013bka} and M{\scriptsize AD}G{\scriptsize RAPH}5\_{\scriptsize A}MC@NLO~\cite{Alwall:2014hca}.
The dashed line represents the cross section which is generated by only the s-channel diagrams,
while the real line shows the cross section of both the t-channel and s-channel diagrams under the following kinematical cuts (for the Vector Boson Fusion (VBF) cuts, see Ref.~\cite{cite:W_pair_fusion}.),
\bal
m_{jj}>500\ \mathrm{GeV}, \hspace{10pt}
|\Delta \eta | > 2.5,
\label{eq:VBFcut}
\eal
where $m_{jj}$ is the invariant mass of the two jets, and $|\Delta \eta|$ is the difference of the pseudo-rapidity of the jets.
In both the cross sections, the basic kinematical cuts~\cite{cite:W_pair_fusion},
\bal
& p_T^j > 30\ \mathrm{GeV}, \hspace{10pt} |\eta_j| < 5.0,
 \hspace{10pt}  p_T^{\ell^{(\prime)}} > 20\ \mathrm{GeV}, \hspace{10pt} |\eta_{\ell^{(\prime)}} | < 2.5,
\label{eq:Basic_cut}
\eal
are taken into account,
where $p_T^j$ and $\eta_j$ are the transverse momentum and the pseudo-rapidity of the jets, and $p_T^{\ell^{(\prime)}}$ and $\eta_{\ell^{(\prime)}}$ are  the transverse momentum and the pseudo-rapidity of $\ell^{(\prime) +}$.
Obviously, the cross section with the VBF cuts is larger than that of s-channel diagrams.
In the following, we use the VBF cuts to obtain the signal events.

We consider the following SM background processes,
\bal
\label{eq:SMBG1}
pp&\rightarrow Z^{(\ast)} (\text{or }\gamma^\ast)jj \rightarrow \ell^+\ell^-jj, \\
\label{eq:SMBG2}
pp&\rightarrow \ell^+\ell^\prime{}^+\nu_\ell\nu_{\ell^\prime}jj. 
\eal
When we investigate the LNV process where the lepton flavor is conserved, we have to consider the first background process in Eq.~(\ref{eq:SMBG1}). 
The number of the SM background events can be reduced by the transverse momentum cut, $p_T^{\ell^{(\prime)}} > 500\ \mathrm{GeV}$, and also multiplying the charge misidentification rate. 
The second process in Eq.~(\ref{eq:SMBG2}) has the missing transverse momentum, and it can be reduced by cut, $\cancel{p}_T < 20\ \mathrm{GeV}$~\cite{Atre:2009rg}, where $\cancel{p}_T$ is the missing transverse momentum.
In Table~\ref{Table:signal_and_BG}, we show the cross sections of the LNV signal $pp\to\mu^+\mu^+jj$ at $\sqrt{s}=14\ \mathrm{TeV}$ and the cross section of the SM backgrounds for each step of the kinematical cuts. In the numerical evaluation, FEYNRULES 2.0~\cite{Alloul:2013bka} and M{\scriptsize AD}G{\scriptsize RAPH}5\_{\scriptsize A}MC@NLO~\cite{Alwall:2014hca} are used.

\renewcommand{\arraystretch}{1.2}
\begin{table}[h]
\begin{center}
\begin{tabular}{ c| c| c| c| c| c|}
 & Basic cut & $+\, \mathrm{VBF\ cut}$ & $+\, \cancel{p}_T\ \mathrm{cut}$ & $+\, p_T^\ell\ \mathrm{cut}$ \\ \hline\hline
 \begin{tabular}{c}
 Signal (pb) \\
 eff. \\
 \end{tabular}
 &
 \begin{tabular}{c}
  4.69 \\
   - \\
  \end{tabular}
 &
 \begin{tabular}{c}
  4.5 \\
   96 \% \\
  \end{tabular}
 &
 \begin{tabular}{c}
 4.5 \\
  100 \% \\
  \end{tabular}
&
\begin{tabular}{c}
2.9 \\
 64 \% \\
 \end{tabular}
 \\ \hline
 \begin{tabular}{c}
$\mu^+\mu^-jj$ (pb) \\
eff. \\
\end{tabular}
&
\begin{tabular}{c}
 $117$ \\
 - \\
 \end{tabular}
 & 
 \begin{tabular}{c}
 $4.1$ \\
 $3.5 \%$ \\
 \end{tabular}
 &
  \begin{tabular}{c}
 $4.1$ \\
 $100 \%$ \\
 \end{tabular}
 &
 \begin{tabular}{c}
 $5.3 \times 10^{-5}$ \\
 $1.3 \times 10^{-3}\  \%$ \\
 \end{tabular}
    \\ \hline
\begin{tabular}{c}
$\mu^+\mu^+\nu_\mu\nu_\mu jj$ (pb)\;  \\
eff. \\
\end{tabular}
& 
\begin{tabular}{c}
$3.71 \times 10^{-3}$ \\
 - \\
 \end{tabular}
 &
 \begin{tabular}{c}
  $1.40 \times 10^{-3}$ \\
  38 \% \\
  \end{tabular}
  & 
   \begin{tabular}{c}
  $6.5 \times 10^{-5}$ \\
  4.6 \% \\
  \end{tabular}
 & 
\begin{tabular}{c}
$6 \times 10^{-9}$ \\
$0.01\  \%$ \\
\end{tabular}
 \\ \hline
\end{tabular}
\caption{ Cross sections of the signal and each background and the efficiency of each kinematical cut. We calculate the signal cross section with the condition, $C_{\mu\mu}^R / \Lambda = 1\ \mathrm{TeV^{-1}}$. The scattering cross sections are calculated with M{\scriptsize AD}G{\scriptsize RAPH}5\_{\scriptsize A}MC@NLO~\cite{Alwall:2014hca}.  }
\label{Table:signal_and_BG}
\end{center}
\end{table}
\renewcommand{\arraystretch}{1.0}

In the following, we consider how the coupling constants $|C^R_{\ell \ell^\prime}/\Lambda|$ can be constrained by searching for the LNV processes $pp\rightarrow \ell^+\ell^\prime{}^+jj $ at the future HL-LHC experiment~\cite{HL-LHC}.
First, we consider the LNV process where the anti-leptons in the final states have the same lepton flavor, $pp\rightarrow \ell^+\ell^+jj $.
The beam energy, $\sqrt{s} = 14\ \mathrm{TeV}$, is much higher than the masses of charged leptons, so that the cross section is insensitive to the flavor of anti-leptons in the final state.
Therefore, we here only consider the process $pp\rightarrow \mu^+\mu^+jj $ and the constraint on $C_{\mu\mu}^R/\Lambda$.
We expect that the constraints on the coupling constants with other flavors, $C_{ee}^R/\Lambda$ and $C_{\tau \tau}^R/\Lambda$, are almost the same as that on $C_{\mu\mu}^R/\Lambda$.

By using the results in Table~\ref{Table:signal_and_BG},
we can estimate the number of the SM background events at the HL-LHC experiment as in Table~\ref{table:number_of_SM_Background}.
The rate of the charge misidentification is assumed to be $1\%$
because we use the kinematical cut $p_\mathrm{T}^\ell > 500\ \mathrm{GeV}$, so that anti-muons have large transverse momenta~\cite{Chatrchyan:2009ae}.
Expected numbers of the background events are respectively $\mathcal{O}(1)$ or much less than $1$ for the processes in Eqs.~(\ref{eq:SMBG1}) and~(\ref{eq:SMBG2}).
Therefore, if we obtain $\mathcal{O}(10)$ events of the LNV signals, 
we can say that they are not from the SM background events but from the signal events via the $\mu^+ \mu^+ W^- W^-$ operator.
\renewcommand{\arraystretch}{1.5}
\begin{table}[h]
\begin{center}
\begin{tabular}{c|c|c|}
 & $\ \mu^+ \mu^- j j\  $ & $\ \mu^+ \mu^+ \nu_{\mu} \nu_{\mu}jj \ $  \\ \hline 
\# of events & $1.6$ & $1.8 \times 10^{-2}$  \\ \hline
\end{tabular}
\caption{The expected number of SM background events at the HL-LHC experiment (with the collision energy of $\sqrt{s} = 14\ \mathrm{TeV}$ and the integrated luminosity of $L = 3000\ \mathrm{fb^{-1}}$). We assume that the rate of charge misidentification is $1\%$.}
\label{table:number_of_SM_Background}
\end{center}
\end{table}
\renewcommand{\arraystretch}{1.0}

\begin{figure}[h]
\begin{center}
\includegraphics[width = 100mm]{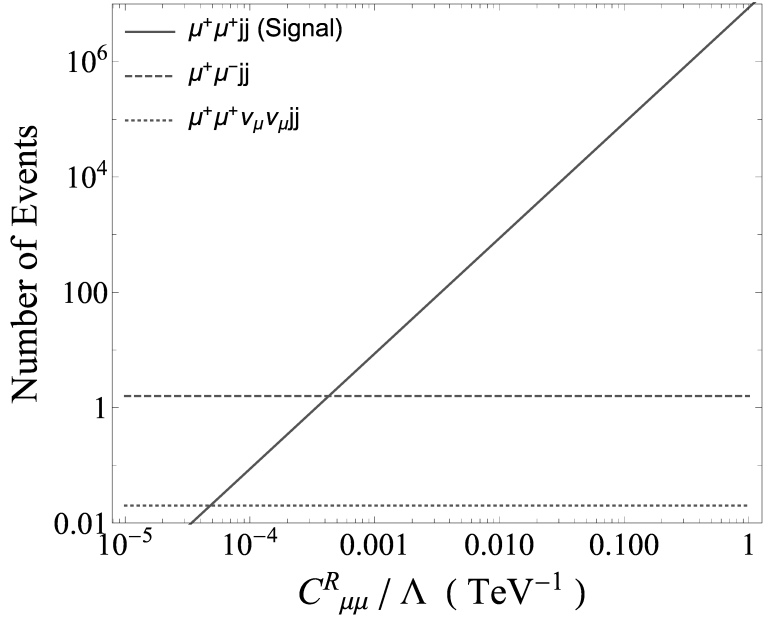}
\caption{ The $C^R_{\mu\mu}/\Lambda$ dependence of the number of $pp\to e^+e^+jj$ events and the numbers of the background events at the HL-LHC  (with the collision energy of $\sqrt{s} = 14\ \mathrm{TeV}$ and the integrated luminosity of $L = 3000\ \mathrm{fb^{-1}}$). It is assumed that the rate of the charge misidentification is $1\%$. }
\label{fig:CuttedSignal_mmjj}
\end{center}
\end{figure}

In Fig.~\ref{fig:CuttedSignal_mmjj}, we show numbers of the signal events and those of the background events at the HL-LHC experiment as a function of $|C_{\mu\mu}^R/\Lambda|$. We assume that the rate of the charge misidentification is $1\%$~\cite{Chatrchyan:2009ae}.
The real line represents numbers of the signal event. The dashed and dotted lines represent numbers of the background events from 
$\mu^+\mu^-jj$ and $\mu^+\mu^+\nu_\mu\nu_\mu jj$, respectively. 
Expected number of the signal event is $\mathcal{O}(10)$
at the point where $|C^R_{\mu\mu}/\Lambda| = 1\, \times 10^{-3} \ \mathrm{TeV^{-1}}$.
Therefore, the LNV event $\mu^+ \mu^+ jj$ is expected to be observed in the region $|C_{\mu\mu}^R/\Lambda| \gtrsim 10^{-3}\ \mathrm{TeV^{-1}}$.
In other words, if we do not have the signal event at the HL-LHC experiment with integrated luminosity $3000\ \mathrm{fb^{-1}}$, we obtain the constraints
$|C_{\mu\mu}^R/\Lambda| \lesssim 10^{-3}\ \mathrm{TeV^{-1}}$.
The similar constraints on $|C_{ee}^R/\Lambda|$ and $|C_{\tau \tau}^R/\Lambda|$ can be obtained if no excess is observed.
Expected upper bounds on the coupling constants of the $\ell^\pm \ell^\pm W^\mp W^\mp$ operators are then given at the HL-LHC
\bal
\left| \frac{ C_{\ell \ell}^R }{ \Lambda } \right| \lesssim 10^{-3}\ \mathrm{TeV^{-1}}.
\eal

\begin{figure}[h]
\begin{center}
\includegraphics[width = 100mm]{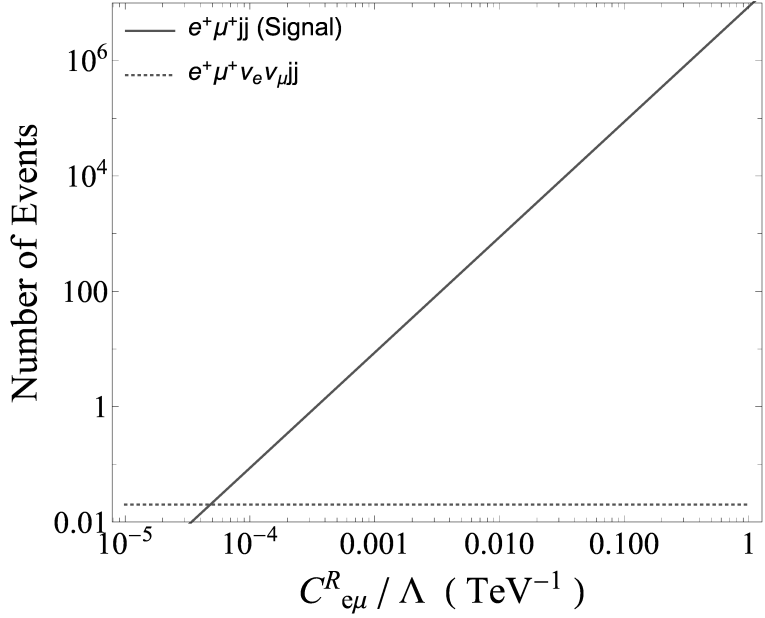}
\caption{ The $C^R_{e\mu}/\Lambda$ dependence of the number of $pp\to e^+\mu^+jj$ event and the number of the background events at Hl-LHC (with the collision energy of $\sqrt{s} = 14\ \mathrm{TeV}$ and the integrated luminosity of $L = 3000\ \mathrm{fb^{-1}}$).}
\label{fig:CuttedSignal_emjj}
\end{center}
\end{figure}

Next, we consider LNV processes where the anti-leptons have different lepton flavor $pp\to\ell^+\ell^\prime{}^+jj\; (\ell \neq \ell^\prime)$.
As the beam energy is much higher than the masses of charged leptons, we only discuss the process $pp\to e^+\mu^+jj$ and the constraint on $C_{e \mu}^R/\Lambda$.
It is expected that constraints on the other coupling constants, $C_{\ell \ell^\prime}^R/\Lambda\; (\ell \neq \ell^\prime)$, are similar to that on $C_{e \mu}^R/\Lambda$.
The most important SM background is $pp\to e^+\mu^+\nu_e\nu_\mu jj$,
and the number of this event is much less than $1$ at the HL-LHC, as shown in Table~\ref{table:number_of_SM_Background}.
In Fig.~\ref{fig:CuttedSignal_emjj}, we show the $|C^R_{e\mu}/\Lambda|$ dependence of the number of the signal event and that of the background event. We assume that the rate of the charge misidentification is $1\%$.
The real line represents the number of the signal event while the dotted line does that of the background event. 
Expected numbers of the signal event is $\mathcal{O}(1)$
at the point where $|C^R_{e\mu}/\Lambda| = 5\, \times 10^{-4} \ \mathrm{TeV^{-1}}$.
Expected numbers of the background event is much less than $1$, and we can observe the LNV event $e^+ \mu^+ jj$ in the region $|C_{e\mu}^R/\Lambda| \gtrsim 5\ \times 10^{-4}\ \mathrm{TeV^{-1}}$.
In other words, if we do not have the signal events at the HL-LHC, we obtain the constraint
$|C_{e\mu}^R/\Lambda| \lesssim 5\ \times 10^{-4}\ \mathrm{TeV^{-1}}$.
For the other set of flavor in $|C_{\ell \ell^\prime}^R/\Lambda|\; (\ell \neq \ell^\prime)$, similar constraints are expected to be obtained.
Expected upper bounds for the coupling constants of the $\ell^\pm \ell^\prime{}^\pm W^\mp W^\mp$ operators $(\ell \neq \ell^\prime)$ are then given at the HL-LHC
\bal
\left| \frac{ C_{\ell \ell^\prime }^R }{ \Lambda } \right| \lesssim 5 \times 10^{-4}\ \mathrm{TeV^{-1}}.
\eal

Consequently,
the expected upper limits on $|C_{\ell \ell^\prime}^R/\Lambda|$ at the HL-LHC are 
\bal
\left| \frac{ C_{\ell \ell}^R }{ \Lambda } \right| \lesssim\ & 1 \times 10^{-3}\ \mathrm{TeV^{-1}},
\\
\left| \frac{ C_{\ell \ell^\prime}^R }{ \Lambda } \right| \lesssim\ & 5 \times 10^{-4}\ \mathrm{TeV^{-1}}\hspace{10pt} (\ell \neq \ell^\prime).
\eal
By using Eq.~(\ref{eq:def_CR}), we can translate thsese upper limits to those on the original coupling constants $C_{\ell  \ell^\prime}^{(9)}/\Lambda_\mathrm{LNV}^5$:
\bal
\left| \frac{ C_{\ell \ell}^{(9)} }{ \Lambda_\mathrm{LNV}^5 } \right| \lesssim\ & 5.5\ \mathrm{TeV^{-5}},
\\
\left| \frac{ C_{\ell \ell^\prime}^{(9)} }{ \Lambda_\mathrm{LNV}^5 } \right| \lesssim\ & 2.8\ \mathrm{TeV^{-5}}\hspace{10pt} (\ell \neq \ell^\prime).
\eal

With the assumption $|C_{\ell \ell^\prime}^{(9)}| = 1$, we can obtain the expected upper limit on the new physics scale from HL-LHC as
\bal
\label{eq:lower_limit_Coll_RH1}
\Lambda_\mathrm{LNV} \gtrsim\  0.71\ \mathrm{TeV}\hspace{10pt} (\ell = \ell^\prime), \\
\label{eq:lower_limit_Coll_RH2}
\Lambda_\mathrm{LNV} \gtrsim\ 0.82  \ \mathrm{TeV} \hspace{10pt} (\ell \neq \ell^\prime). 
\eal
These bound mean that if the processes $pp\to\ell^+\ell^\prime{}^+jj$ are not observed  
at the HL-LHC, the new physics scale $\Lambda_\mathrm{LNV}$ is higher than $0.71$ or $0.82\ \mathrm{TeV}$.\footnote{
We focus on the final states with $\ell^+$s whose transverse momenta are larger than $500\ \mathrm{GeV}$. Therefore the typical energy scale for the subprocess involving the effective operator might be larger than the lower bounds in Eqs.~(\ref{eq:lower_limit_Coll_RH1}) and~(\ref{eq:lower_limit_Coll_RH2}).
In such a case, the EFT approach would not be appropriate any more, 
and the lower bounds given in Eqs.~(\ref{eq:lower_limit_Coll_RH1}) and~(\ref{eq:lower_limit_Coll_RH2}) might not give reliable bounds on LNV new physics which couples right-handed charged leptons. 
}
Conversely, they mean that if the scale of new physics is lower than $0.71$ or $0.82\ \mathrm{TeV}$, the LNV signal $pp\to\ell^+\ell^\prime{}^+jj$ via \llww operator can be observed at the HL-LHC. 
Then direct signal from new particles which cause \llww operator may also be observed at the HL-LHC depending on the origin of the LNV. 

If the integrated luminosity is higher than $3000\ \mathrm{fb^{-1}}$ at future $pp$ collider experiments with the same energy, the expected lower limit is also higher than those in Eqs.~(\ref{eq:lower_limit_Coll_RH1}) and~(\ref{eq:lower_limit_Coll_RH2}). 
Then, the situation where the signal of $pp\to\ell^+\ell^\prime{}^+jj$ via \llww operator would be observed without any direct evidences of new particles may happen, depending on the scale of the LNV.

\subsection{ The constraint on $C^L_{\ell \ell^\prime}/\Lambda$}
\begin{figure}[h]
\begin{center}
\includegraphics[width=170mm]{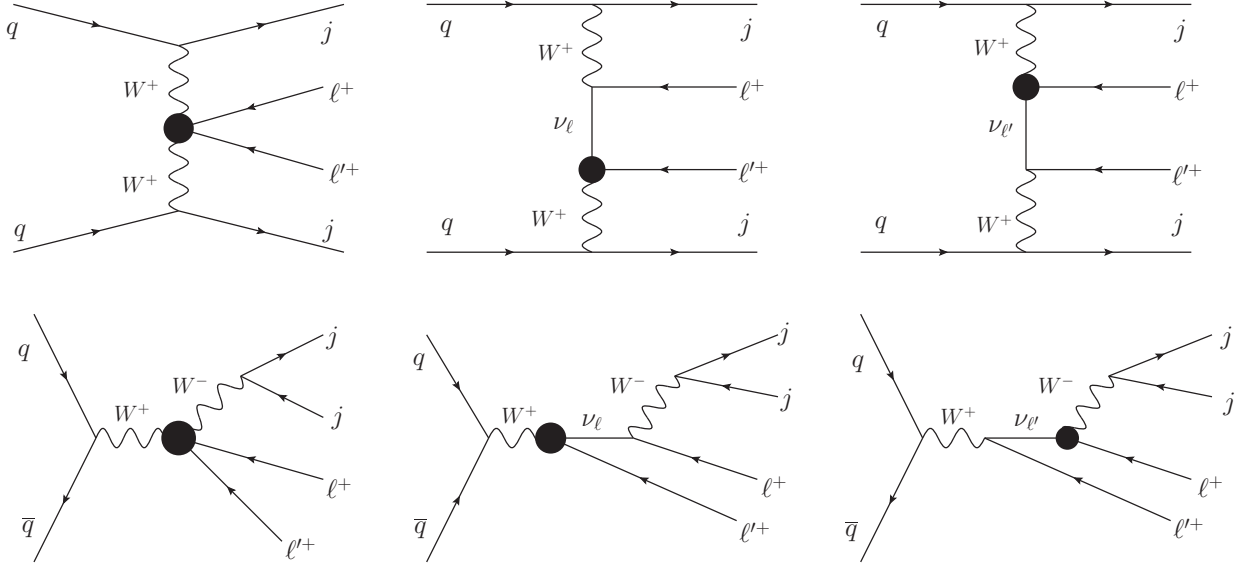}
\caption{Feynman diagrams for the process $pp\rightarrow \ell^+\ell^\prime{}^+jj $ via the dimension-seven LNV operators.
In the case $\ell \neq \ell^\prime$, there are additional diagrams with exchange $\ell \leftrightarrow \ell^\prime$.}
\label{fig:lljjLH}
\end{center}
\end{figure}
In this section, we investigate processes $pp\rightarrow \ell^+\ell^\prime{}^+jj$ with left-handed charged leptons, which are generated by the dimension-seven LNV operators in Eq.~(\ref{eq:dim7LNV}). 
These processes are represented by the Feynman diagrams in Fig.~\ref{fig:lljjLH}.
As compared to the cases with right-handed charged leptons which are generated by the dimension-nine LNV operators in Eq.~(\ref{eq:dim9LNV}), additional diagrams are generated by the three-point vertices in Eq.~(\ref{eq:dim7LNV_expanded}).
However, the contribution from these diagrams is negligibly small.
As a result, constraints on $C_{\ell \ell^\prime}^L/\Lambda$ are almost the same as those on $C_{\ell \ell^\prime}^R/\Lambda$,
and we can estimate upper bounds on $C_{\ell \ell^\prime}^L/\Lambda$ at the HL-LHC as
\bal
\left| \frac{ C_{\ell \ell}^L }{ \Lambda } \right| \lesssim\ & 1 \times 10^{-3}\ \mathrm{TeV^{-1}},
\\
\left| \frac{ C_{\ell \ell^\prime}^L }{ \Lambda } \right| \lesssim\ & 5 \times 10^{-4}\ \mathrm{TeV^{-1}}\hspace{10pt} (\ell \neq \ell^\prime).
\eal
By using Eq.~(\ref{eq:def_CL}), we can translate thsese upper limits to those on the symmetric part of the original coupling constants $(C_{\ell  \ell^\prime}^{(7)}+C_{\ell  \ell^\prime}^{(7)})/\Lambda_\mathrm{LNV}^3$:
\bal
\left| \frac{ C_{\ell \ell}^{(7)} }{ \Lambda_\mathrm{LNV}^3 } \right| \lesssim\ & 0.17\ \mathrm{TeV^{-3}},
\\
\left| \frac{ C_{\ell  \ell^\prime}^{(7)}+C_{\ell  \ell^\prime}^{(7)} }{ 2\Lambda_\mathrm{LNV}^3 } \right| \lesssim\ & 8.3 \times 10^{-2}\ \mathrm{TeV^{-3}}\hspace{10pt} (\ell \neq \ell^\prime).
\eal
With assumption $|C_{\ell  \ell^\prime}^{(7)}+C_{\ell  \ell^\prime}^{(7)}|/2 = 1$, we can obtain the expected upper limit on the new physics scale from HL-LHC as
\bal
\Lambda_\mathrm{LNV} \gtrsim\  1.8\ \mathrm{TeV}\hspace{10pt} (\ell = \ell^\prime), \\
\Lambda_\mathrm{LNV} \gtrsim\ 2.3 \ \mathrm{TeV} \hspace{10pt} (\ell \neq \ell^\prime). 
\eal

Both for the left-handed and right-handed $e^\pm e^\pm W^\mp W^\mp$ operators, the constraint from the current $0\nu\beta\beta$ data is more stringent than expected constraint from the HL-LHC. However, the HL-LHC can be useful to test other \llww operators with the other set of flavor of charged leptons.

\section{Conclusion}
\label{sec:Conclusion}

We have investigated phenomenological consequences of the \llww operators. These operators can contain important information for the origin of tiny neutrino masses which is independent of that from the Weinberg operator. 
We have obtained constraints on the coefficients of the \llww operators 
by the neutrino oscillation data. 
Upper bounds on the coefficients have also been examined by using the data for LNV processes such as neutrinoless double beta decays and the $\mu^-$-$e^+$ conversion.
In addition, we have found that the \llww operators can be directly tested by 
searching for the LNV processes via the same sign W boson fusion process at the HL-LHC.
By the combination of these current and future experiments, we can access dimension-seven and dimension-nine LNV operators in the gauge invariant effective field theory and can further deeply understand the origin of tiny neutrino masses.

\begin{acknowledgments}
M.~A.~was supported in part by Japan Society for the
Promotion of Science (JSPS), Grant-in-Aid for Scientific Research (Grant
No. 17K05412).
K.~E.~was supported in part by the Sasakawa Scientific Research Grant from The Japan Science Society.
S.~K.~was supported in part
by Grant-in-Aid for Scientific Research on Innovative Areas,
the Ministry of Education, Culture, Sports, Science and Technology,
No.~16H06492 and No.~18H04587, and also
by JSPS, Grant-in-Aid for Scientific Research (Grant
No. 18F18022 and No. 18F18321).
\end{acknowledgments}

\newpage

\newpage
\appendix
\section{ Models where dimension seven LNV operators are yielded }
\label{appendix:dim-7_model}
\begin{figure}[b]
\begin{center}
\includegraphics[width=150mm]{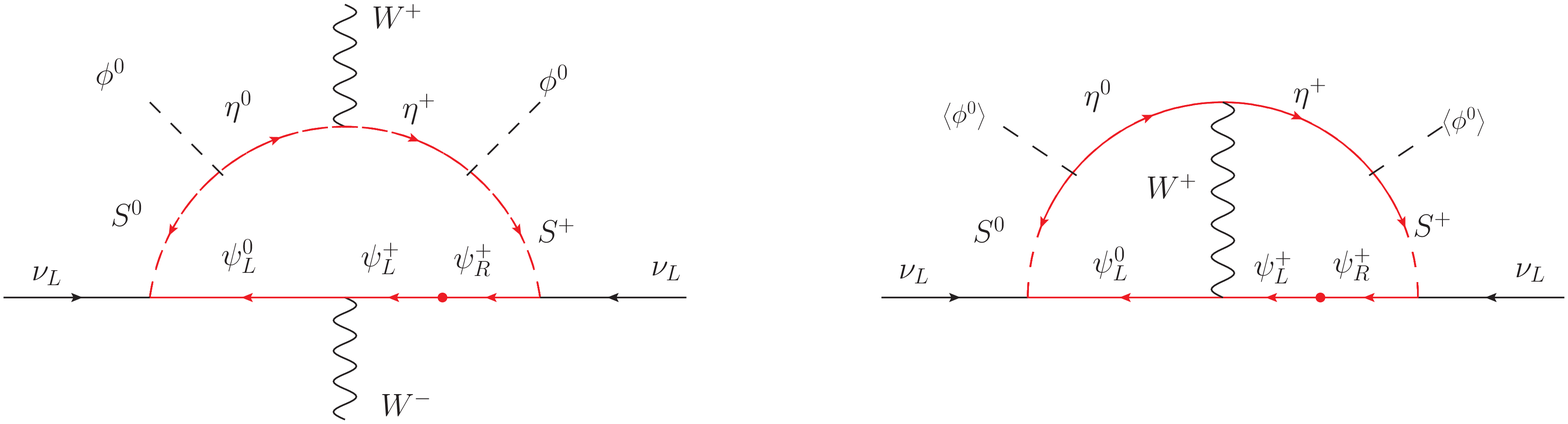}
\caption{Realization of the dimension-seven operators and neutrino masses in Model-I.}
\label{fig:Dim-7_model}
\vspace{50pt}
\includegraphics[width=150mm]{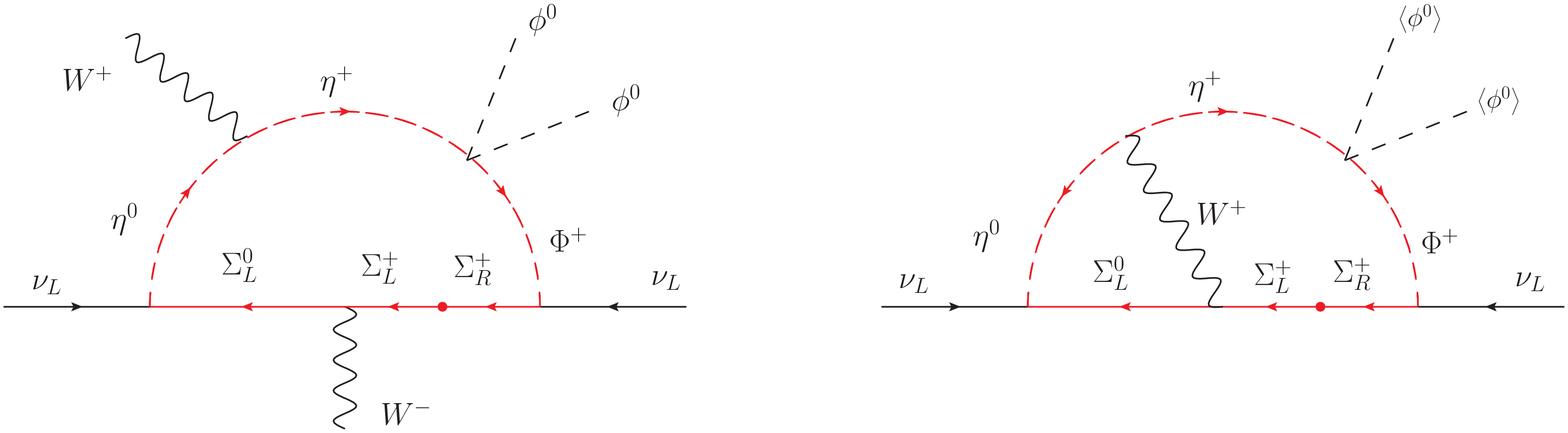}
\caption{Realization of the dimension-seven operators and neutrino masses in  Model-II.}
\label{fig:Dim-7_model2}
\end{center}
\end{figure}

We here show two models where the dimension-seven LNV operators,
\bal
\frac{ C^{ ( 7 ) }_{\ell \ell^\prime} }{ \Lambda_\mathrm{LNV}^3 }
\left( \, \overline{ \tilde{ L }_{ \ell } } D_\mu L_{ \ell^\prime } \right)
\left( \tilde{ \phi}^\dagger D^\mu \phi \right)
+ \mathrm{ h.c. },
\eal
are yielded and neutrino masses are generated by the Feynman diagram in Figs.~\ref{fig:Dim-7_model} and \ref{fig:Dim-7_model2}.

In the first model (Model-I), fields in Table~\ref{table: New fields in the first model} are added to the SM.
\begin{table}[b]
\begin{tabular}{cc}
\begin{minipage}[b]{0.45\hsize}
	\begin{center}
	\begin{tabular}{c|c|c|c|c|}
	 & $\psi_a$ & $\eta$ & $S^+$ & $S^0$ \\ \hline
	 Spin & $1/2$ & \multicolumn{3}{c|}{ 0 } \\ \hline
	 $\mathrm{ SU(2)_L }$ & ${\bf 2}$ & ${\bf 2}$ & ${\bf 1}$ & ${\bf 1}$ \\ \hline
	 $\mathrm{ U(1)_Y }$ & $1/2$ & $1/2$ & $1$ & $0$ \\ \hline
	 $\mathrm{ U(1)^\prime }$ & $q$ & $q$ & $q$ & $-q$ \\ \hline
	 \end{tabular}
	 \caption{Additional fields in Model-I}
	 \label{table: New fields in the first model}
	 \end{center}
\end{minipage} & \hspace{10pt}
\begin{minipage}[b]{0.45\hsize}
	\begin{center}
	\begin{tabular}{c|c|c|c|}
	 & $\Sigma_a$ & $\Phi$ & $\eta$ \\ \hline
	 Spin & $1/2$ & \multicolumn{2}{c|}{0} \\ \hline
	 $\mathrm{ SU(2)_L }$ & ${\bf 3}$ & ${\bf 2}$ & ${\bf 2}$  \\ \hline
	 $\mathrm{ U(1)_Y }$ & $1$ & $3/2$ & $1/2$  \\ \hline
	 $\mathrm{ U(1)^\prime }$ & $q$ & $q$ & $q$  \\ \hline
	 \end{tabular}
	 \caption{Additional fields in Model-II}
	 \label{table: New fields in the second model}
	\end{center}
\end{minipage}
\end{tabular}
 \end{table}  
The model has a new global symmetry $\mathrm{U(1)^\prime}$.
It is an unbroken symmetry after the electroweak symmetry breaking, and we expect that the model can explain a dark matter problem too.
The new fermions $\psi_a = ( \psi_a^+,\, \psi_a^0 )^\mathrm{T}\ ( a=1,2,3)$ are vector-like $\mathrm{SU(2)_L}$ doublets, and they have Dirac mass terms $m_{\psi_a} \overline{\psi_a}\psi_a$.
The model has three kinds of new scalar fields $\eta$, $S^+$ and $S^0$. 
One of the new scalar fields $\eta = ( \eta^+,\, \eta^0 )^\mathrm{T}$ 
is a $\mathrm{SU(2)_L}$ doublet.
Other scalars $S^+$ and $S^0$ are $\mathrm{SU(2)_L}$ singlets. 
All of new scalars do not obtain the vacuum expectation value.
After the electroweak symmetry breaking, the charged scalars $\eta^+$ and $S^+$ and the neutral scalars $\eta^0$ and $S^0$ are mixed via the three-point scalar interactions,
\bal
\kappa_1(\phi^\dagger \eta) S^0 
+ \kappa_2 ( \phi^\dagger \tilde{\eta} ) S^+
+ \mathrm{h.c.},
\eal
where $\kappa_1$ and $\kappa_2$ are coupling constants, and $\phi$ is the Higgs field and $\tilde{\eta} = i \sigma_2 \eta$. 

The model has the following new Yukawa interactions.
\bal
(Y_R)_{ \ell a }\, \overline{ L_\ell}\, P_R\, \psi_a\, S^- 
+ (Y_L)_{ \ell a }\, \overline{ \tilde{L}_\ell }\, P_L\, \psi_a\, S^0
+ \mathrm{h.c.},
\eal
where the operators $P_R$ and $P_L$ are the chirality projection operators.
Then, the dimension-seven LNV operators are generated via the left Feynman diagram in Fig.~\ref{fig:Dim-7_model}.
By using this operator, Majorana masses of neutrinos are generated via the right Feynman diagram in Fig.~\ref{fig:Dim-7_model}.

The second model (Model-II) has the new fields listed in Table~\ref{table: New fields in the second model}.
This model has a new unbroken global symmetry $\mathrm{U(1)^\prime}$.
The new fermions
\renewcommand{\arraystretch}{1.5}
 \bal
 \Sigma_a
 =
 \left(
 	\begin{array}{cc}
	\frac{ \Sigma_a^0}{ \sqrt{ 2 } } & \ \Sigma_a^{++} \\ 
	\Sigma_a^0 & \ - \frac{ \Sigma_a^0 }{ \sqrt{ 2 } } \\
	\end{array}
\right), \hspace{10pt} ( a=1,2,3 ),
\eal
\renewcommand{\arraystretch}{1.0}
\hspace{-10pt} are vector-like $\mathrm{SU(2)_L}$ triplets, and they have Dirac mass terms $M_{\Sigma_a} \mathrm{Tr}[ \overline{ \Sigma_a } \Sigma_a ]$. 
Both of new scalars $\Phi = ( \Phi^{++},\, \Phi^{+} )$ and $\eta = ( \eta^+,\, \eta^0 )$ are $\mathrm{SU(2)_L}$~doublets.
The singly charged scalar fields $\Phi^+$ and $\eta^+$ are mixed via the four-point scalar interaction,
\bal
\kappa_1^\prime ( \Phi^\dagger \phi ) ( \tilde{ \eta }^\dagger \phi ) + \mathrm{h.c.}
\eal

The dimension-seven operators are generated using the above scalar interaction and the following new Yukawa interactions,
\bal
( Y_L^\prime )_{ \ell a }\, \overline{ L_\ell }\, P_R\, \Sigma_a\, \tilde{\Phi}
+ ( Y_R^\prime )_{\ell a }\, \overline{ \tilde{L}_\ell }\, P_L\, \Sigma_a\, \tilde{\eta}
+ \mathrm{h.c.},
\eal
where $\tilde{\Phi} = i \sigma_2 \Phi$. 
Then, the dimension-seven LNV operators are generated via the left Feynman diagram in Fig.~\ref{fig:Dim-7_model2}.
By using this operator, Majorana masses of neutrinos are generated via the right Feynman diagram in Fig.~\ref{fig:Dim-7_model2}.
Detailed discussions of these models are beyond the scope of this paper, which will be given elsewhere~\cite{cite:future}

\section{Renormalization of higher-dimensional LNV oeprators}
\label{appendix:Renormalization of two point functions of neutrinos}

We here show details of the renormalization procedure used for the calculation in Sec.~\ref{sec:Neutrino masses}.
First, we discuss the renormalization of two-point functions of neutrinos which are generated by dimension-seven operators in Eq.~(\ref{eq:dim7LNV}).
We use the following renormalized operators,
\bal
\mathcal{L}_\mathrm{tree}
=
&\, \frac{1}{2}\, \overline{\nu_{a}}\, \Bigl(\, i\cancel{\partial} - m_{\nu_a}\, \Bigr)\, \nu_{a}
+ \frac{ e }{ \sqrt{ 2 } s_\mathrm{w} }\, 
	\Bigl(
		U_{\ell a}^\ast\, \overline{ \nu_{a} }\, \gamma^\mu\, P_L\, \ell \, W_{\mu}^+
		+ U_{\ell a}\, \overline{ \ell }\, \gamma^\mu\, P_L\, \nu_{a}\, W_\mu^-
	\Bigr)
\nonumber \\
& + \Biggl[
		-\frac{ i e }{ 2\sqrt{2} s_\mathrm{w} } 
		\frac{ v^2 }{ \Lambda_\mathrm{LNV}^3 }
		C^{ ( 7 ) }_{\ell \ell^\prime}
			\Bigl\{
				 (\, 
	  			U_{\ell^\prime a }\, \overline{ \ell^c } \, \partial_\mu  \, P_L\, \nu_{a } 
	 			 - U_{\ell a}\, \overline{ \nu_{a}^c }\, \partial_\mu \, P_L\, \ell^\prime
				\, ) W^{+\mu}
\nonumber \\
& \hspace{135pt}
				+ \frac{ i e }{ \sqrt{2} s_\mathrm{w} } U_{\ell a}\, U_{\ell^\prime b}\, 
				\overline{ \nu_a }\, P_L\, \nu_{ b }\, W_{\mu}^-\, W^{+ \mu}
			\Bigr\} 
	+ \mathrm{h.c.}
	\Biggr]
\nonumber \\
& - \frac{ v^2 }{ 2 } 
	\Bigl(
		F_{\ell \ell^\prime}^{(7)} \, U_{\ell b}^\ast\, U_{\ell^\prime a}^\ast\, 
		\overline{ \nu_{b} }\,  P_R\, \partial^2\, \nu_{a}
		+ F_{\ell \ell^\prime}^{(7)\ast} \, U_{\ell b}\, U_{\ell^\prime a}\, 
		\overline{ \nu_{a} }\,  P_L\, \partial^2\, \nu_{b}
	\Bigr),
\label{eq:renormalized_lagrangian_dim7}
\eal
where $U$ is the PMNS matrix, and neutrino fields are in the mass eigenstate basis
$\nu_{\ell,L} = U_{\ell a}\, \nu_{a,L}$, ($\ell = e, \mu, \tau$ and $a = 1,2,3$).
The Majorana neutrino fields $\nu_{a}$ are defined as in Eqs.~(\ref{eq:Majorana_neutrino1}) and~(\ref{eq:Majorana_neutrino2}) such that they satisfy Majorana conditions $\nu_a^c = \nu_a$.
The mass matrix of Majorana neutrinos $m_{\nu_a}$ is generated by the Weinberg operator as
\bal
m_{\nu_a} \delta_{ab} = \frac{ v^2 }{ \Lambda_5 }\, U_{\ell a}\, C_{\ell \ell^\prime}^{(5)}\, U_{\ell^\prime b},
\label{eq:Neutrino_mass_Weinberg}
\eal
where $\delta_{ab}$ is the Kronecker Delta.
The LNV operators proportional to $C_{\ell \ell^\prime}^{(7)}$ and their hermitian conjugations are generated by the dimension-seven operators in Eq.~(\ref{eq:dim7LNV}) which are the origin of the \llww operators with left-handed charged leptons.
The LNV operators in the last line of Eq.~(\ref{eq:renormalized_lagrangian_dim7})
are generated by the operators in Eq.~(\ref{eq:dim7counter}).
Their counter terms are used to eliminate logarithmic divergences proportional to the squared momentum of the external neutrino.
Counter terms which are needed to eliminate divergences in two-point functions of neutrinos are given by
\bal
\mathcal{L}_\mathrm{counter}
= &\, 
\frac{ i }{ 2 } \delta^{1} Z_{ab}\, \overline{ \nu_{a} }\, \cancel{\partial}\, P_L\, \nu_{b}
+ \frac{ 1 }{ 2 } \delta^{1} m_{ab}\, \overline{\nu_{a} }\,P_L\, \nu_{b}
+ \frac{ 1 }{ 2 } \delta^{1} F_{ab}\, \overline{\nu_{a} }\, P_L\, \partial^2\, \nu_{b}
+ \mathrm{h.c.},
\label{eq:counter_terms_dim7}
\eal
where $\delta^{1} Z_{ab}$, $ \delta^{1} m_{ab}$ and $\delta^{1} F_{ab}$ are $\mathcal{O}(\hbar)$ coefficients of counter terms.
The coefficients $\delta^{1} Z_{ab}$ and $ \delta^{1} m_{ab}$
satisfy the following conditions;
\bal
&\delta^{1} Z_{ab} = (\delta^{1} Z_{ba})^\ast \\
&\delta^{1} m_{ab}= \delta^{1} m_{ba} 
\eal

In the 'tHooft-Feynman gauge, the renormalized amputated two-point functions for neutrinos in the mass eigenstate basis $i\Sigma_{ab}(\cancel{p})$ are given by
\bal
\label{eq:def_SimgaL:App}
& i \Sigma_{ab}(\cancel{p}) = \, i \Sigma^L_{ab}(\cancel{p})\, P_L + \, i \Bigl(\Sigma^L_{ab}(\cancel{p})\Bigr)^\ast\, P_R,
\eal
where
\bal
& \Sigma^L_{ab}(\cancel{p}) =  \, \widetilde{\Sigma}^L_{ab}(p^2)\, +\, \delta^1 \widetilde{Z}_{ab} \, \cancel{p}
							 +\, \delta^1 \tilde{m}_{ab}
							 +\, \delta^1 \widetilde{F}_{ab}\, p^2,
\\
&\widetilde{\Sigma}^L_{ab}(p^2) = 
		- \frac{ v^2 }{ 4 \Lambda_\mathrm{LNV}^3 } \left( \frac{e }{s_\mathrm{w} } \right)^2
		\nonumber \\
		& \hspace{ 60pt }\times
			\Biggl\{
				\int_k\, \frac{ 4 }{ k^2 + m_W^2 }\, 
				\Bigl( (U^\mathrm{T} C^{(7)} U)_{ab} + ( U^\mathrm{T} C^{(7)} U)_{ba} \Bigr)
				\nonumber \\
				& \hspace{70pt}
				+ \int_0^1\ \mathrm{d}x\, \int_k\, \frac{ 1 }{ ( k^2 + \Delta_\ell )^2 } 
				\left( - \frac{ k^2 }{ d } + x^2 p^2 \right)
				 \Bigl(
				 	 (U^\mathrm{T} C^{(7)} )_{a\ell}U_{\ell b} + ( a\leftrightarrow b) 
				\Bigr)\, U_{\ell b} 
				\nonumber \\
				& \hspace{70pt}
				+ \int_0^1\ \mathrm{d}x\, \int_k\, \frac{ 1 }{ ( k^2 + \Delta_\ell )^2 } 
				\, xp^2\, 
				 \Bigl(
				 	 ( C^{(7)} U )_{\ell a}U_{\ell b} + ( a\leftrightarrow b) 
				\Bigr)\, U_{\ell b}
			\Biggr\},
\\
& \Delta_\ell = (1-x)m_{\ell}^2 + x m_W^2 - x (1-x) p^2,
\\
& \delta^1 \widetilde{Z}_{ab} =  \, \frac{ 1 }{ 2 }\, ( \delta^1 Z_{ba} + \delta^1 Z_{ab}^\ast ),
\\
& \delta^1 \tilde{m}_{ab} =  \, \frac{ 1 }{ 2 }\, ( \delta^1m_{ab} + \delta^1 m_{ba} ),
\\
& \delta^1 \widetilde{F}_{ab} =  \, \frac{ 1 }{ 2 }\, ( \delta^1 F_{ab} + \delta^1 F_{ba} ),
\eal
and $\int_k = \int\, \mathrm{d}^dk / (2\pi)^d$ represents the integral over all $d$-dimensional euclidean momentum space.
We impose the following on-shell conditions~\cite{Aoki:1982ed,Grimus:2016hmw};
\bal
&m_{\nu_b} \delta^1 \widetilde{Z}_{ba}
=
- \widetilde{\Sigma}_{ab}^L(m_{\nu_b}^2 ) 
- m_{\nu_b}^2\, \delta^1 \widetilde{F}_{ab}
- \delta^1 \tilde{m}_{ab}. 
\\
& 
\delta^1 \widetilde{Z}_{aa}
 =
- 2 \mathrm{Re} \left[
				\delta^1 \widetilde{F}_{ab}
				+
				\left.
				\frac{ \mathrm{d} \widetilde{\Sigma}^L_{aa} }{ \mathrm{d} p^2 }
				\right|_{p^2 = m_{\nu_a}^2}
				\right].
\eal
We cannot determine all coefficients of the counter terms with only imposing the on-shell conditions,
so that we impose the additional condition,
\bal
\left. \frac{ \mathrm{d} \widetilde{\Sigma}_{ab}^L }{ \mathrm{d} p^2 } \right|_{p^2 = 0}
= 0.
\eal
Then, $\Sigma_{ab}^L(\cancel{p})$ in Eq.~(\ref{eq:def_SimgaL:App}) are given by
\bal
\Sigma^L_{aa}(\cancel{p}) = & - 2 m_{\nu_a } (\cancel{p} - m_{\nu_a} )
\mathrm{Re}\Bigl[
				A^\prime_{aa}(m_{\nu_a}^2) - A^\prime_{aa}(0)
				+ B_{aa}(m_{\nu_a}^2) - B_{aa}(0)
				+ m_{\nu_a}^2\, B_{aa}^\prime(m_{\nu_a}^2)
			\Bigr]
\nonumber \\
&+  A_{aa}(p^2) + p^2\, B_{aa}(p^2) - A_{aa}(m_{\nu_a}^2) - m_{\nu_a}^2\, B_{aa}(m_{\nu_a}^2)
\nonumber \\
&
- ( p^2 - m_{\nu_a}^2 )\Bigl( A_{aa}^\prime (0) + B_{aa}(0) \Bigr),
\\
\Sigma_{ab}^L(\cancel{p})
= &
\, \frac{ \cancel{p} - m_{\nu_a} }{ m_{\nu_b}^2 - m_{\nu_a}^2 }
\Biggl\{
		m_{\nu_a} \Bigl( A_{ab}(m_{\nu_a}^2) - A_{ab}(m_{\nu_b}^2)
						+ m_{\nu_a}^2 B_{ab}(m_{\nu_a}^2)
						- m_{\nu_b}^2 B_{ab}(m_{\nu_b}^2)
\nonumber \\
						& \hspace{90pt}
						+ (m_{\nu_b}^2 - m_{\nu_a}^2 )
						\Bigl( A^\prime_{ab}(0) + B_{ab}(0) \Bigr)
					\Bigr)
\nonumber \\
						& \hspace{60pt}
		+ m_{\nu_b} \Bigl( A_{ab}^\ast(m_{\nu_a}^2) - A_{ab}^\ast(m_{\nu_b}^2)
						+ m_{\nu_a}^2 B_{ab}^\ast(m_{\nu_a}^2)
						- m_{\nu_b}^2 B_{ab}^\ast(m_{\nu_b}^2)
\nonumber \\
						& \hspace{90pt}
						+ (m_{\nu_b}^2 - m_{\nu_a}^2 )
						\Bigl( A^{\prime\ast}_{ab}(0) + B^\ast_{ab}(0) \Bigr)
					\Bigr)
\Biggr\}
\nonumber \\
& +  A_{ab}(p^2) + p^2\, B_{ab}(p^2) - A_{ab}(m_{\nu_a}^2) - m_{\nu_a}^2\, B_{ab}(m_{\nu_a}^2)
\nonumber \\
&
- (p^2 - m_{\nu_a}^2 )\Bigl( A_{ab}^\prime (0) + B_{ab}(0) \Bigr),
\hspace{10pt} ( a\neq b),
\eal
where
\bal
&A_{ab}(p^2) 
=
 \, \frac{ 1}{ 32\pi^2} \frac{ m_W^2 }{ \Lambda_\mathrm{LNV}^3 }
\Bigl( (U^\mathrm{T} C^{(7)} )_{a\ell} U_{\ell b} + (a \leftrightarrow b) \Bigr)
\int_0^1 \mathrm{d}x\, \Bigl( (1-x)m_{\ell}^2 + x m_W^2 \Bigr) \ln \Delta_\ell,
\\
&B_{ab}(p^2) 
=\, 
\frac{1}{32 \pi^2} \frac{ m_W^2 }{ \Lambda_\mathrm{LNV}^3 }
\Biggl\{
	\Bigl( (U^\mathrm{T} C^{(7)} )_{a\ell} U_{\ell b} + ( a \leftrightarrow b) \Bigr)
	\int_0^1 \mathrm{d}x\, x(1+x)\ln \Delta_\ell
\nonumber \\
& \hspace{100pt}
+ \Bigl( (C^{(7)} U )_{\ell a} U_{\ell b} + ( a \leftrightarrow b) \Bigr)
	\int_0^1 \mathrm{d}x\, x \ln \Delta_\ell
\Biggr\},
\\
& A^\prime_{ab}(p^2) = \, \frac{ \mathrm{d} A_{ab} }{ \mathrm{d} p^2 },
\\
& B^\prime_{ab}(p^2) = \, \frac{ \mathrm{d} B_{ab} }{ \mathrm{d} p^2 }.
\eal
Masses of charged leptons $m_{\ell}$ and those of neutrinos $m_{\nu_a}$ are smaller than that of the weak bosons $m_W$,
so that the leading term of $\Sigma_{ab}^{L}(\cancel{p})$ is given by
\bal
\label{eq:SigmaL1loopLH}
\Sigma_{ab}^L(\cancel{p})
\simeq &
\ - \frac{1}{ 16\pi^2 }\, U_{\ell a} \frac{ C_{ \ell \ell^\prime }^L }{ \Lambda }\, U_{\ell^\prime b}
\, f\left(\frac{p^2}{m_W^2}\right),
\eal
where
\bal
f(x) = & \ \frac{ 1 }{ 36 x^2 } \Bigl( x(6+57x-97x^2) + 6(1-x)^2 (11x+1) \ln(1-x) \Bigr).
\eal
and Eq.~(\ref{eq:def_CL}) is used.
The formula in Eq.~(\ref{eq:SigmaL1loopLH}) are that in Eq.~(\ref{eq:prediction_SigmaL}) in Sec.~\ref{sec:Neutrino masses}.

Next, we show the renormalization of two-point functions of neutrinos which are generated by dimension-nine operators in Eq.~(\ref{eq:dim9LNV}).
We use the following renormalized operators;
\bal
\mathcal{L}_\mathrm{tree}
=
&\, \frac{1}{2}\, \overline{\nu_{a}}\, \Bigl(\, i\cancel{\partial} - m_{\nu_a}\, \Bigr)\, \nu_{a}
+ \frac{ e }{ \sqrt{ 2 } s_\mathrm{w} }\, 
	\Bigl(
		U_{\ell a}^\ast\, \overline{ \nu_{a} }\, \gamma^\mu\, P_L\, \ell \, W_{\mu}^+
		+ U_{\ell a}\, \overline{ \ell }\, \gamma^\mu\, P_L\, \nu_{a}\, W_\mu^-
	\Bigr)
\nonumber \\
& + \Biggl[
		- \frac{ e^2 }{ 8 s_\mathrm{w}^2 } 
		\, \frac{ v^4 }{ \Lambda_\mathrm{LNV}^5 }\, 
		\, C_{\ell \ell^\prime}^{(9)}\, 
		\overline{ \ell^c }\, P_R\, \ell^\prime\, W_\mu^+\, W^{+\mu}
		+
		\frac{ i v^2 }{ 4 }\, F_{\ell \ell^\prime}^{\prime(7)}\, U_{\ell^\prime a}^\ast
		\overline{\ell}\,  \gamma^\mu\, P_R\, \nu_{a}\, W^-_{\mu}
		+ \mathrm{h.c.}
	\Biggr],
\label{eq:renormalized_lagrangian_dim9}
\eal
where the mass matrix of Majorana neutrinos $m_{\nu_a}$ is generated by the Weinberg operator as in Eq.~(\ref{eq:Neutrino_mass_Weinberg}).
The LNV operators proportional to $C_{\ell \ell^\prime}^{(9)}$ and their hermitian conjugations are generated by the dimension-nine operators in Eq.~(\ref{eq:dim9LNV})
, which are the origin of the \llww operators with right-handed leptons.
The LNV operators proportional to $F_{\ell \ell^\prime}^{\prime (7)}$ are generated by the operators in Eq.~(\ref{eq:dim7counter2}).
The LNV operators from Eq.~(\ref{eq:dim9LNV}) generate Majorana masses of neutrinos at two-loop level, while those from Eq.~(\ref{eq:dim7counter2}) do Majorana masses of neutrinos at one-loop level.
Counter terms which are needed to eliminate divergences in two-point functions of neutrinos are given by
\bal
\mathcal{L}_\mathrm{counter}
= &\, 
\frac{ i }{ 2 } \Bigl( \delta^{1} Z_{ab} + \delta^{2} Z_{ab} \Bigr)\, \overline{ \nu_{a} }\, \cancel{\partial}\, P_L\, \nu_{b}
+ \frac{ 1 }{ 2 } \Bigl( \delta^{1} m_{ab} + \delta^{2} m_{ab} \Bigr)\, \overline{\nu_{a} }\,P_L\, \nu_{b}
\nonumber \\
&+ \frac{ 1 }{ 2 } \delta^{1} F^\prime_{a\ell} \, \overline{\nu_{a} }\, \gamma ^\mu\, P_L\, \ell\, W^+_\mu
+ \mathrm{h.c.},
\label{eq:counter_terms_dim9}
\eal
where $\delta^{1} Z_{ab}$, $\delta^{1} m_{ab}$ and $\delta^{1} F^\prime_{a\ell}$ are $\mathcal{O}(\hbar)$, and $\delta^{2} Z_{ab}$ and $ \delta^{2} m_{ab}$ are $\mathcal{O}(\hbar^2)$ coefficients of counter terms.
The coefficients $\delta^{1} Z_{ab}$, $\delta^{1} m_{ab}$, $\delta^{2} Z_{ab}$ and $ \delta^{2} m_{ab}$ satisfy the relations
\bal
& \delta^{1} Z_{ab} = (\delta^{1} Z_{ba})^\ast, \hspace{10pt}
 \delta^{2} Z_{ab} = (\delta^{2} Z_{ba})^\ast, \\
& \delta^{1} m_{ab} = \delta^{1} m_{ba}, \hspace{10pt}
 \delta^{2} m_{ab} = \delta^{2} m_{ba}.
\eal

At one-loop level, two-point functions of neutrinos are generated via the operators which are proportional to $F_{\ell \ell^\prime}^{\prime (7)}$.
The Feynman diagrams are shown in Fig.~\ref{fig:MassDiagram_for_Renormalization}.
In the 'tHooft-Feynman gauge, the renormalized amputated two-point functions at one-loop level $i\Sigma^{1-loop}_{ab}(\cancel{p})$ are given by
\bal
& i\Sigma^{1-loop}_{ab}(\cancel{p})
=  \, i\Sigma^{L, 1-loop}_{ab}(\cancel{p})\, P_L + \, i \Bigl( \Sigma^{L, 1-loop}_{ab}(\cancel{p}) \Bigr)^\ast\, P_R,
\eal
where
\bal
\label{eq:SigmaL1loop}
& \Sigma^{L, 1-loop}_{ab}(\cancel{p})
= 
\widetilde{\Sigma}^{L,1-loop}_{ab}(p^2)  + \delta^1 \tilde{m}_{ab} + \delta^1 \widetilde{Z}_{ab}\, \cancel{p},
\\
& \widetilde{\Sigma}^{L,1-loop}_{ab}(p^2)
= 
- \frac{ i v^3 }{ \sqrt{ 2 } } \left( \frac{ e }{ s_\mathrm{w} } \right)^2
\Bigl\{
	(U^\mathrm{T} F^{\prime (7)} )_{a\ell} m_\ell U_{\ell b} + (a \leftrightarrow b)
\Bigr\}
\nonumber \\
& \hspace{80pt} \times  \int_0^1 \mathrm{d}x\, \int_k\, \frac{ 1 }{ ( k^2 + \Delta_\ell )^2 },
\\
& \Delta_{\ell} = (1-x) m_\ell^2 + x m_W^2 - x(1-x)p^2, \\
& \delta^1 \widetilde{Z}_{ab} =  \, \frac{ 1 }{ 2 }\, ( \delta^1 Z_{ba} + \delta^1 Z_{ab}^\ast ),
\\
& \delta^1 \tilde{m}_{ab} =  \, \frac{ 1 }{ 2 }\, ( \delta^1m_{ab} + \delta^1 m_{ba} ).
\eal
In order to determine the coefficients of counter terms $\delta^1 \tilde{m}_{ab}$ and $\delta^1 \widetilde{Z}_{ab}$, we impose the following on-shell conditions~\cite{Aoki:1982ed,Grimus:2016hmw};
\bal
&m_{\nu_b} \delta^1 \widetilde{Z}_{ba}
=
- \widetilde{\Sigma}_{ab}^{L,1-loop}(m_{\nu_b}^2 ) 
- \delta^1 \tilde{m}_{ab}. 
\\
& 
\delta^1 \widetilde{Z}_{aa}
 =
- 2 \mathrm{Re} \left[
				\left.
				\frac{ \mathrm{d} \widetilde{\Sigma}^{L,1-loop}_{aa} }{ \mathrm{d} p^2 }
				\right|_{p^2 = m_{\nu_a}^2}
				\right].
\eal
Then, $\Sigma^{L,1-loop}_{ab}(\cancel{p})$ in Eq.~(\ref{eq:SigmaL1loop}) are given by
\bal
\Sigma_{aa}^{L,1-loop}
= &
- 2 m_{\nu_a} (\cancel{p} - m_{\nu_a} )\mathrm{Re} 
\left[ \left. \frac{ \mathrm{d} G_{aa} }{ \mathrm{d} p^2 } \right|_{p^2 =m_{\nu_a}^2 } \right]
 + G_{aa}(p^2) - G_{aa}(m_{\nu_a}^2),
\\
\Sigma_{ab}^{L,1-loop}
=&
\frac{ \cancel{p}-m_{\nu_a} }{ m_{\nu_b}^2 - m_{\nu_a}^2 }
	\Bigl\{
	m_{\nu_a} \Bigl( G_{ab}(m_{\nu_a}^2) - G_{ab}(m_{\nu_b}^2) \Bigr)
	+ m_{\nu_b} \Bigl( G_{ab}^\ast(m_{\nu_a}^2) - G_{ab}^\ast(m_{\nu_b}^2) \Bigr)
	\Bigr\}
	\nonumber \\
	&
	 + G_{ab}(p^2) - G_{ab}(m_{\nu_a}^2), \hspace{10pt} ( a\neq b),
\eal
where
\bal
G_{ab}(p^2) = &\frac{ i v^3 }{ 32 \pi^2 } \left( \frac{ e }{ s_\mathrm{w} } \right)^2
\Bigl\{
	( U^\mathrm{T} F^{\prime (7)} )_{b\ell} m_\ell U_{\ell a} + ( a \leftrightarrow b) 
\Bigr\}
\int_0^1\mathrm{d}x\, \ln\Delta_\ell.
\eal

\begin{figure}[h]
\begin{center}
\includegraphics[width=100mm]{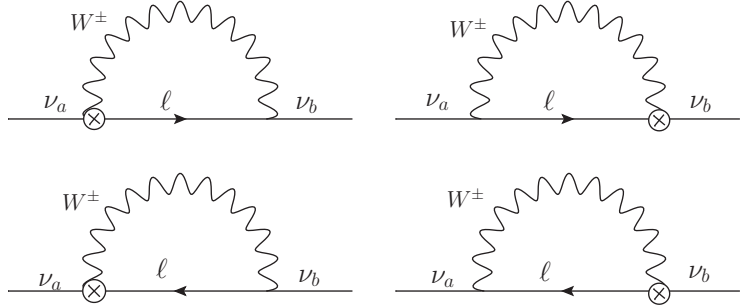}
\caption{ Feynman diagrams for two-point functions of neutrinos via the counter term of the dimension-seven operators in Eq.~(\ref{eq:dim7counter2}).}
\label{fig:MassDiagram_for_Renormalization_Appendix}
\end{center}
\end{figure}

At two-loop level, two-point functions of neutrinos are generated via the dimension-nine LNV operators in Eq.~(\ref{eq:dim9LNV}).
The Feynman diagrams are shown in Fig.~\ref{fig:MassRH}.
In the 'tHooft-Feynman gauge, the renormalized amputated two-point functions at two-loop level $i\Sigma_{ab}^{2-loop}(\cancel{p})$ are given by
\bal
 i\Sigma^{2-loop}_{ab}(\cancel{p})
= & \, i\Sigma^{L, 2-loop}_{ab}(\cancel{p})\, P_L 
+ \, i \Bigl( \Sigma^{L, 2-loop}_{ab}(\cancel{p}) \Bigr)^\ast\, P_R,
\eal
where
\bal
\label{eq:SigmaL2loop}
 \Sigma^{L, 2-loop}_{ab}(\cancel{p})
= &
\widetilde{\Sigma}^{L,2-loop}_{ab}(p^2)  + \delta^2 \tilde{m}_{ab} + \delta^2 \widetilde{Z}_{ab}\, \cancel{p} + \delta^1 \widetilde{F}^\prime_{ab,\ell} I_\ell(p^2),
\\
 \widetilde{\Sigma}^{L,2-loop}_{ab}(p^2)
= &
\frac{v^4}{ 8 \Lambda_\mathrm{LNV}^5 } \left( \frac{ e }{ s_\mathrm{w} }\right)^4
\Bigl( C_{\ell \ell^\prime }^{(9)} + C_{\ell^\prime \ell }^{(9)} \Bigr) \, U_{\ell a} \, U_{\ell^\prime b}\, m_\ell\, m_{\ell^\prime}
I_\ell(p^2)\, I_{\ell^\prime}(p^2)
\eal
with
\bal
I_\ell (p^2) 
= &
\int_0^1 \int_k \frac{ 1 }{ ( k^2 +\Delta_\ell )^2 },
\\
\Delta_{\ell} = & (1-x) m_\ell^2 + x m_W^2 - x(1-x)p^2, \\
 \delta^2 \widetilde{Z}_{ab} = &  \, \frac{ 1 }{ 2 }\, ( \delta^2 Z_{ba} + \delta^2 Z_{ab}^\ast ),
\\
\delta^2 \tilde{m}_{ab} =&  \, \frac{ 1 }{ 2 }\, ( \delta^2 m_{ab} + \delta^2 m_{ba} ),
\\
\delta^1 \widetilde{F}^\prime_{ab,\ell} 
= &
- 2\sqrt{2} \left( \frac{ e }{ s_\mathrm{w} } \right)
\left\{
	\delta^1 F^\prime_{b\ell} m_\ell U_{\ell a} + i \delta^1 F_{a \ell}^\prime m_\ell U_{\ell b}
\right\}. 
\eal
The term proportional to $\delta^1 \widetilde{F}^{\, \prime}_{ab,\ell}$ comes from Feynman diagrams in Fig.~\ref{fig:MassDiagram_for_Renormalization_Appendix}, which are generated via the counter term proportional to $\delta^1 F^{\, \prime}_{a\ell}$.
We here assume that coefficients of the other counter terms, for example coefficients for wave function renormalization of the weak bosons or charged leptons, are zero
because they do not need to eliminate divergences.
\vspace{20pt}

We impose the following on-shell conditions~\cite{Aoki:1982ed,Grimus:2016hmw};
\bal
m_{\nu_b} \delta^2 \widetilde{Z}_{ba}
=&
- \widetilde{\Sigma}_{ab}^{L,2-loop} (m_{\nu_b}^2) - \delta^2 \tilde{m}_{ab} 
- \delta^1 \widetilde{F}_{ab,\ell}\, I_{\ell}(m_{\nu_b}^2),
\\
\delta^2 \widetilde{Z}_{aa}
=&
- 2 m_{\nu_a}\, 
\mathrm{Re}
		\left[
			\left.\left(
			\frac{ \mathrm{d} \widetilde{\Sigma}_{aa}^{L,2-loop} }{ \mathrm{d} p^2 }
			+ \delta^1 \widetilde{F}^\prime_{ab,\ell}\, 
			\frac{ \mathrm{d} I_\ell }{ \mathrm{d} p^2 }
			\right)\right|_{p^2 = m_{\nu_a}^2}
		\right].
\eal
We cannot determine all coefficients of counter terms with only the on-shell conditions, so that
we impose the additional condition,
\bal
\left.
\frac{ \mathrm{d} }{ \mathrm{d} p^2 } 
\left( \widetilde{\Sigma}_{ab}^{L,2-loop}(p^2) + \delta^1 F_{ab,\ell}^\prime\, I_\ell(p^2) \right)
\right|_{p^2 =0}
=
0.
\eal
Then, $\Sigma_{ab}^{L,2-loop}(\cancel{p})$ in Eq.~(\ref{eq:SigmaL2loop}) are given by
\bal
\Sigma_{aa}^{L,2-loop}(\cancel{p})
= & - 2 m_{\nu_a}( \cancel{p} - m_{\nu_a} )\, \mathrm{Re}
\Biggl[
	A_{aa}^{\ell \ell^\prime}\, I^\prime_\ell (m_{\nu_a}^2)
	\Bigl( I_{\ell^\prime}(m_{\nu_a}^2) - I_{\ell^\prime}(0) \Bigr)
\Biggr]
\nonumber \\
&+ A_{aa}^{\ell \ell^\prime}\, \Bigl( H_{\ell \ell^\prime}(p^2) - H_{\ell \ell^\prime}(m_{\nu_a^2})
\Bigr),
\\
\Sigma_{ab}^{L,2-loop}(\cancel{p})
=&
\frac{ \cancel{p} - m_{\nu_a} }{ m_{\nu_b}^2 - m_{\nu_a}^2 }
\Biggl[
		A_{ab}^{\ell \ell^\prime}\, m_{\nu_a}\, 
		\Bigl( H_{\ell \ell^\prime}(m_{\nu_a}^2) - H_{\ell \ell^\prime}(m_{\nu_b}^2) \Bigr)
		\nonumber \\
		&\hspace{70pt}
		+ (A_{ab}^{\ell \ell^\prime})^\ast\, m_{\nu_b}\, 
		\Bigl( H_{\ell \ell^\prime}(m_{\nu_a}^2) - H_{\ell \ell^\prime}(m_{\nu_b}^2) \Bigr)^\ast
\Biggr]
\nonumber \\
&+ A_{ab}^{\ell \ell^\prime}\, \Bigl( H_{\ell \ell^\prime}(p^2) - H_{\ell \ell^\prime}(m_{\nu_a}^2) \Bigr), \hspace{10pt} ( a\neq b),
\eal
where
\bal
A_{ab}^{\ell \ell^\prime} = & \frac{ v^4 }{ 8 \Lambda_\mathrm{LNV}^5 } \left( \frac{ e }{ s_\mathrm{w} } \right)^4\, \Bigl( C_{ \ell \ell^\prime }^{(9)} + C_{ \ell^\prime \ell }^{(9)}  \Bigr)
\, U_{\ell a}\, U_{\ell^\prime b}\, m_\ell\, m_{\ell^\prime},
\\
H_{\ell \ell^\prime}(p^2) = & ( I_{\ell}(p^2) - I_{\ell}(0) ) ( I_{\ell^\prime}(p^2) - I_{\ell^\prime}(0) ),
\\
I^\prime_{\ell}(p^2) = \, & \frac{ \mathrm{d} I_{\ell} }{ \mathrm{d} p^2 }.
\eal
Masses of charged leptons $m_{\ell}$ and neutrinos $m_{\nu_a}$ are smaller than that of the weak bosons $m_W$,
so that the leading terms of $\Sigma_{ab}^{1-loop}(\cancel{p})$ and  $\Sigma_{ab}^{2-loop}(\cancel{p})$ are give by
\bal
\label{eq:RHSigma1-loop}
\Sigma_{ab}^{L, 1-loop}(\cancel{p})
\simeq &
 - \frac{ i v^3 }{ 32 \pi^2 } \left( \frac{ e }{ s_\mathrm{w} } \right)^2
 \Bigl\{
 	(U^\mathrm{T} F^{\prime(7)})_{b\ell} \, m_\ell \, U_{\ell a} 
	+ (U^\mathrm{T} F^{\prime(7)})_{a \ell} \, m_\ell \, U_{\ell b} )
\Bigr\}
\, g\left( \frac{ p^2 }{ m_W^2 } \right),
\\
\label{eq:RHSigma2-loop}
\Sigma_{ab}^{L, 2-loop}(\cancel{p})
\simeq & - \frac{ 1 }{ 128\pi^4 } \left( \frac{ e }{ s_\mathrm{w} } \right)^2
\, \frac{ C_{\ell \ell^\prime}^R }{\Lambda}\, U_{\ell a}\, U_{\ell^\prime b}\, 
m_\ell\, m_{\ell^\prime}\, \biggl\{ g\left( \frac{ p^2 }{ m_W^2 } \right) \biggr\}^2,
\eal
where
\bal
g(x) = &\, 1 + \frac{ (1-x) }{ x }\, \ln(1-x),
\eal
and Eq.~(\ref{eq:def_CR}) is used.
The sum of Eq.~(\ref{eq:RHSigma1-loop}) and Eq.~(\ref{eq:RHSigma2-loop}) is
Eq.~(\ref{eq:predictionSigmaLRH}) in Sec.~\ref{sec:Neutrino masses}.

\end{document}